\documentclass[journal]{IEEEtran}
\usepackage{graphicx}
\usepackage{subfig}
\usepackage{amsmath} 
\usepackage{epsfig}
\usepackage{breqn}
\usepackage{bm}
\usepackage{amssymb}
\usepackage{mathtools}

\usepackage{color}
\usepackage{cite}
\usepackage{balance}
\usepackage{empheq}
\newcommand{\bb}{\bm}
\usepackage{algorithm,algorithmic}
\newcommand{\txtc}[1]{\textcolor{black}{#1}}
\DeclareMathOperator*{\argminA}{arg\,min} 
\newcommand{\argmaxD}{\arg\!\max} 
\DeclareMathOperator*{\argmaxA}{arg\,max} 
\begin{document}
\title{Direct Reconstruction of Linear Parametric Images from Dynamic PET Using Nonlocal Deep Image Prior}
\author{Kuang Gong,  Ciprian Catana, Jinyi Qi and Quanzheng Li
\thanks{This work was supported by the National Institutes of Health under grants R21AG067422, R03EB030280, RF1AG052653 and P41EB022544.}
\thanks{K.~Gong and Q.~Li are with Gordon Center for Medical Imaging, Massachusetts General Hospital and Harvard Medical School, Boston, MA 02114 USA (e-mail: kgong@mgh.harvard.edu, li.quanzheng@mgh.harvard.edu). }
\thanks{Ciprian Catana is with Martinos Center for Biomedical Imaging, Massachusetts General Hospital and Harvard Medical School, Boston, MA 02114 USA  (e-mail: ccatana@mgh.harvard.edu)}
\thanks{J.~Qi is with the Department of Biomedical Engineering, University of California, Davis, CA 95616 USA  (e-mail: qi@ucdavis.edu)}
}
\begin{twocolumn}

\maketitle 
\begin{abstract}
Direct reconstruction methods have been developed to estimate parametric images directly from the measured PET sinograms by combining the PET imaging model and tracer kinetics in an integrated framework. Due to limited counts received, signal-to-noise-ratio (SNR) and resolution of parametric images produced by direct reconstruction frameworks are still limited. Recently supervised deep learning methods have been successfully applied to medical imaging denoising/reconstruction when large number of high-quality training labels are available. For static PET imaging, high-quality training labels can be acquired by extending the scanning time. However, this is not feasible for dynamic PET imaging, where the scanning time is already long enough. In this work, we proposed an unsupervised deep learning framework for direct parametric reconstruction from dynamic PET, which was tested on the Patlak model and the relative equilibrium Logan model. The training objective function was based on the PET statistical model. The patient’s anatomical prior image, which is readily available from PET/CT or PET/MR scans, was supplied as the network input to provide a manifold constraint, and also utilized to construct a kernel layer to perform non-local feature denoising. The linear kinetic model was embedded in the network structure as a $1 \times 1$ convolution layer.  Evaluations based on dynamic datasets of $^{18}$F-FDG and $^{11}$C-PiB tracers show that the proposed framework can outperform the traditional and the kernel method-based direct reconstruction methods.
\end{abstract}

\begin{IEEEkeywords}
Direct reconstruction, dynamic PET, deep neural network, unsupervised learning, positron emission tomography
\end{IEEEkeywords}

\section{Introduction}

Positron Emission Tomography (PET) is an important imaging modality with essential roles in oncology, neurology and cardiology studies.  {\it{In vivo}} physiology activities inside the tissue can be revealed noninvasively through the injection of specifically designed PET tracers. Compared to the widely employed static PET protocol, dynamic PET acquires multiple time frames and accordingly each voxel/region-of-interest (ROI) has multiple temporal measurements instead of one. The voxel-wise PET parametric map can be derived from the temporal measurements according to a pre-selected kinetic model, which can provide quantitative spatial distribution of metabolism, receptor binding or blood flow. It can achieve better performance than static PET for lesion detection \cite{dimitrakopoulou2001dynamic,yang2020influx}. Due to various physical degradation factors, the image quality of PET is inferior to other imaging modalities. The ill-conditionness of solving kinetic models further challenges PET parametric imaging. All of these compromise the accuracy and potentials of PET parametric imaging for early detection, staging and longitudinal monitoring. Developing advanced processing/reconstruction methods to improve the accuracy of PET parametric imaging is greatly needed.

The conventional way to calculate PET parametric maps is to first reconstruct sequential dynamic images from frame-wise projection data, and then estimate the kinetic parameters based on pixel-wise fitting of the time activity curves (TACs). However, it is difficult to accurately model the noise in the image space through this indirect reconstruction approach. Direct reconstruction methods were proposed to estimate kinetic parameters directly from raw measurement in one step \cite{kamasak2005direct,tsoumpas2008study,wang2009generalized,matthews2010direct,rahmim2012direct,angelis2018direct,petibon2020pet}, and thus generate parametric maps with improved signal-to-noise ratio (SNR) due to better noise modeling. However, due to limited counts received and the  physical degradation factors, further improvement in image quality of direct reconstruction is still desirable.  Various approaches have been proposed to further improve direct PET image reconstruction based on joint-entropy \cite{tang2010direct}, Bowsher prior-based penalty function \cite{loeb2015direct}, dictionary learning \cite{yang2019sparsity} and the kernel method \cite{gong2017direct}.

Deep learning methods have been widely applied to PET image denoising \cite{wang20183d,chen2019ultra,cui2019pet,lu2019investigation,klyuzhin2019use,hashimoto2019dynamic,sanaat2020projection,angelis2020denoising}, reconstruction\cite{gong2018iterative,yang2018artificial,mehranian2020model,lim2020improved}, and direct sinogram-to-image mapping \cite{haggstrom2019deeppet,whiteley2020directpet,kandarpa2020dug,hu2020dpir}. One challenge of applying deep learning to dynamic PET is the lack of high-quality training labels. For static PET, training labels can be obtained by prolonging the scan time. However, this is not feasible for dynamic PET, where the scan time is already too long. To address this training-label challenge, an alternative approach is the deep image prior (DIP) proposed by Ulyanov {\it{et al}} based on the observation that convolutional neural networks (CNNs) have the intrinsic ability to regularize a variety of ill-posed inverse problems \cite{ulyanov2017deep}.  Under the original DIP framework, random noise was supplied as the network input and the noisy image itself was used as the training label to generate denoised images. For PET imaging, anatomical priors from Magnetic Resonance (MR) or Computed Tomography (CT) exist and have been proposed to be supplied as the network input to further improve the original DIP framework \cite{gong2018dip,cui2019pet}.

Recently Wang {\it{et al}} proposed the nonlocal neural networks \cite{wang2018non} to improve the video classification accuracy, which was achieved by feature denoising through the nonlocal operation inside the network. In this framework, the nonlocal layer calculation was based on the features extracted from the previous layer, whose function is similar to the attention mechanism. For PET imaging, similar to the kernel method \cite{hutchcroft2016anatomically}, the nonlocal layer can be calculated from the anatomical prior instead of the extracted features, which has lower image noise and higher spatial resolution. It can also reduce the number of trainable parameters and thus reduce the training difficulty, which is essential for unsupervised deep learning. 

In this work, we proposed a novel direct reconstruction framework inspired by the DIP framework and the nonlocal concept. No high-quality training labels were needed in this proposed framework, the patient's anatomical prior image was utilized as the network input, and the final training objective function was formulated based on the Poisson distribution of the dynamic PET sinograms. Two linear kinetic models, the Patlak model \cite{patlak1983graphical} and the Relative Equilibrium (RE) Logan model \cite{zhou2009consistent}, were employed in this study to test the feasibility of the proposed framework. Regarding the network structure, 3D U-net \cite{cciccek20163d} was employed as the backbone and the kinetic model was embedded into the network structure as a kinetic-model layer. Furthermore, a nonlocal layer based on the patient' anatomical prior image was designed to perform feature denoising and facilitate the modeling of long-range pixel dependencies. Regarding the implementation, the alternating direction method of multipliers (ADMM) algorithm \cite{boyd2011distributed} was utilized to optimize the whole objective function and the L-BFGS algorithm \cite{zhu1997algorithm} was employed for the network training subproblem. In addition, for the RE Logan model, a new dynamic-data binning strategy was proposed to preserve the independent and identically distributed (i.i.d.) assumption of the dynamic sinograms.  

The major contributions of this work include: (1) a novel unsupervised deep learning-based direct PET image reconstruction framework was proposed; (2) a specifically designed network structure which includes the kinetic-model layer and the nonlocal layer was developed for the proposed framework; (3) clinical dynamic 18F-FDG and 11C-PiB datasets were utilized to test the feasibility of the proposed framework based on the Patlak and RE Logan models. This paper is organized as follows. Section 2 introduces the related background, the proposed framework and implementation details. Section 3 describes the simulations and real data used in the evaluation. Experimental results are shown in section 4, followed by discussions in section 5. Finally, conclusions are drawn in Section 6.

\section{Methods}
\subsection{Direct PET image reconstruction}
Let us denote the unknown dynamic PET images after decay correction as  $\bb{x}\in \mathbb{R}^{N \times T}  = [\bb{x}_1, ..., \bb{x}_T]$ and the measured dynamic data  as $\bb{{y}}  \in \mathbb{R}^{M \times T} = [\bb{y}_1, ..., \bb{y}_T]$,  where $N$, $M$ and $T$ are the numbers of voxels, lines-of-responses (LORs) and dynamic frames, respectively. The image intensity in the $k^{th}$ frame after decay correction, $\bb{x}_{k}\in \mathbb{R}^{N }$, can be expressed as
\begin{equation}
\bb{x}_{k} (\bb{\theta})= \int_{t_{s,k}}^{t_{e,k}} \bb{c}(\tau; \bb{\theta})d\tau, 
\label{eq:tac}
\end{equation}
where $t_{s,k}$ and $t_{e,k}$ are the start time and end time of frame $k$, and  ${\bb{c}}(t; \bb{\theta})$ is the tracer concentration image at time $t$ whose formula is based on the kinetic parameters $\bb{\theta}$ and the chosen kinetic model.

Conventionally, images are reconstructed frame-by-frame and then the kinetic parameters are estimated by fitting the time activity curves to the specific kinetic model. Here we use the direct reconstruction framework, which directly estimate the parametric image $\bb{\theta}$ from the measured dynamic data  $\bb{{y}}$.  The mean of measured dynamic data $\bb{\bar{y}} \in \mathbb{R}^{M \times T}$ can be expressed as  \cite{qi1998high}
\begin{equation}
\bb{\bar{y}}(\bb{\theta}) = \boldsymbol{P}\bb{x}(\bb{\theta})  + \bb{r},
\label{eq:systemmodel}
\end{equation}
where  $\bb{P} \in \mathbb{R}^{M \times N}$ models the radioactive decay, photon attenuation, and detector efficiency as well as the detection-probability and motion-transformation matrices,  and $\bb{r} \in \mathbb{R}^{M \times T}$ represents the expectation of randoms and scatters. The log-likelihood function based on the i.i.d. Poisson-distribution assumption of $\bb{{y}}$  can be written as
\begin{equation}
L(\boldsymbol{y}|\boldsymbol{\theta}) \propto \sum_{k=1}^T\sum_{i=1}^M (\bb{y}_k)_i \log (\bar{\bb{y}}(\bb{\theta})_k)_i - (\bar{\bb{y}}(\bb{\theta})_k)_i.
\label{likelihood}
\end{equation}

\begin{figure}[t]
\centering
\subfloat{\includegraphics[trim=4cm 10.5cm 0.6cm 3.3cm, clip, width=5in]{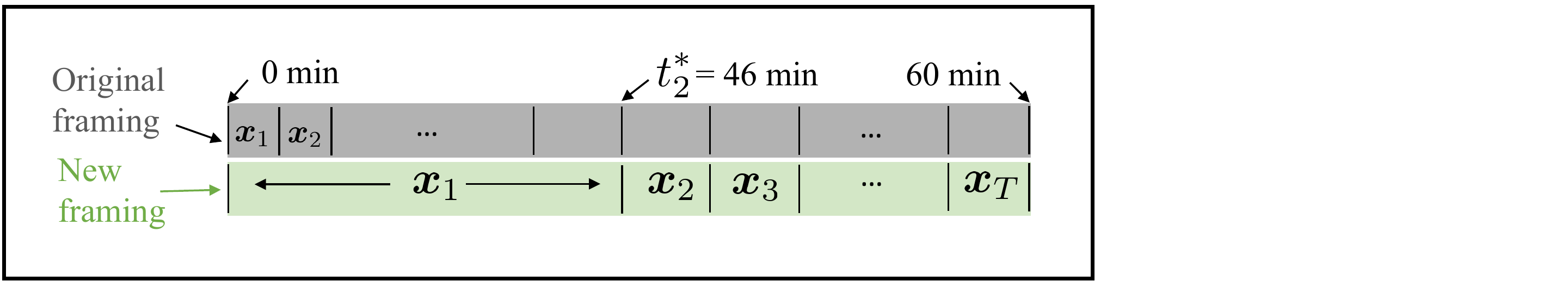}}
\caption{\small{The proposed data binning strategy for the RE Logan model-based direct reconstruction framework. The scan time indicated in the plot is based on the 11C-PiB scanning protocol described in Sec.~\ref{sec:pib_study}. }}
\label{fig:data_bining}
\end{figure}

\begin{figure*}[htp]
\centering
\subfloat{\includegraphics[trim=0.9cm 3cm 0.2cm 2.2cm, clip, width=7in]{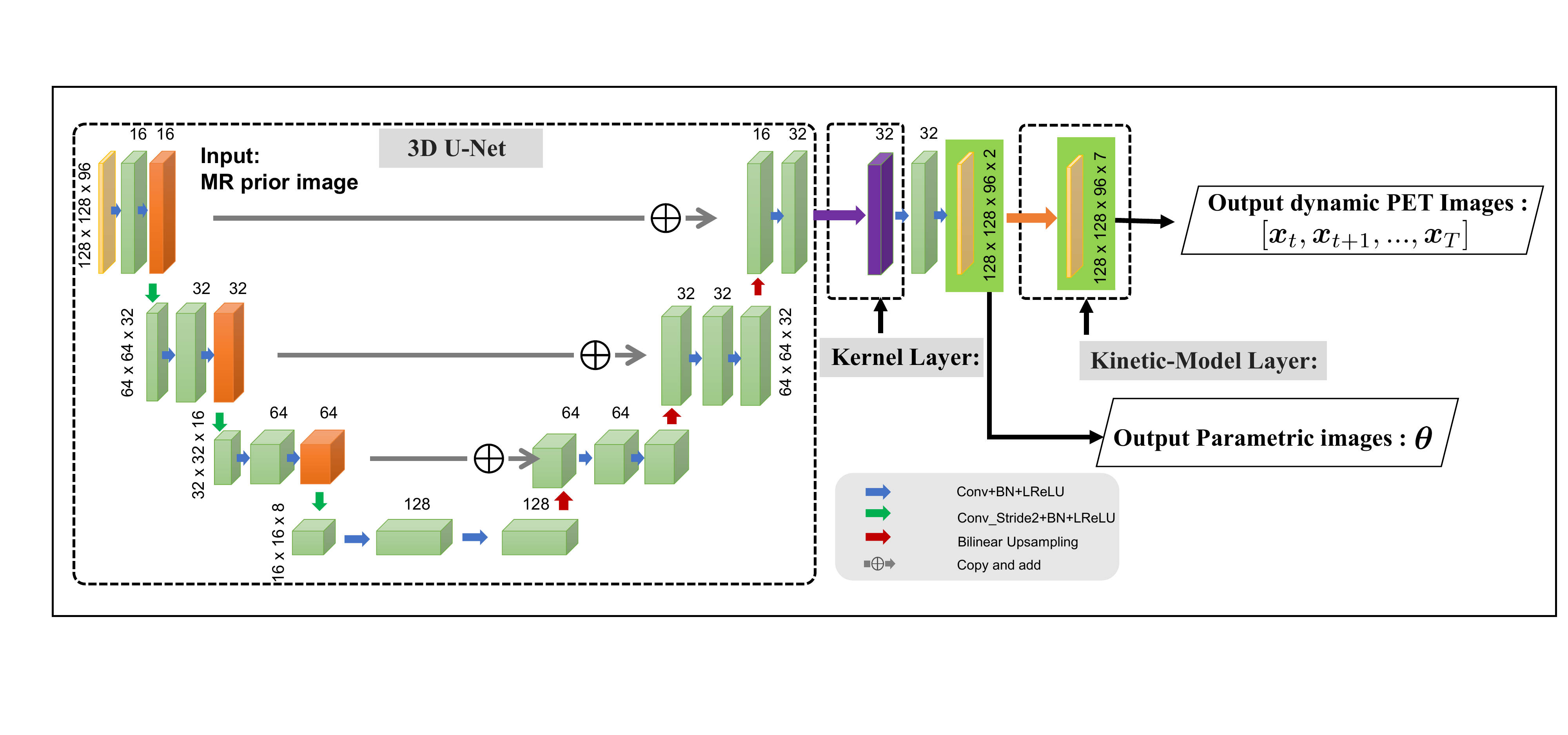}}
\caption{\small{The schematic plot of the proposed network structure. It contains the 3D U-Net as the backbone with the specifically designed kernel layer and kinetic-model layer.  The numbers shown in the plot are based on the simulation study described in Sec.~\ref{sec:simulation}.}}
\label{fig:network_structure}
\end{figure*}

\subsection{ Proposed framework for the Patlak model}\label{sec:patlak}
\subsubsection{Patlak model}
Based on the Patlak model  \cite{patlak1983graphical}, for tracers with at least one irreversible compartment, after reaching a steady time $t^{\ast}$, $\bb{c}(t; \bb{\theta})$ can be approximated as \cite{patlak1983graphical} 
\begin{equation}
{\bb{c}(t; \bb{\theta})} = {\bb{\kappa}}\int_0^t C_p(\tau)d\tau + {\bb{b}}{C_p(t)}, t \geq  t^{\ast},
\label{eq:patlak_model}
\end{equation}
 where $C_p(t)$ is the tracer concentration in the plasma, $\bb{\kappa}\in \mathbb{R}^{N }$ and $\bb{b}\in \mathbb{R}^{N }$ are the Patlak slope and intercept images, respectively. Correspondingly, $\bb{\theta} = [\bb{\kappa}, \bb{b}]$. Embedding equation (\ref{eq:patlak_model}) into (\ref{eq:tac}), $\bb{x}_k$ can be expressed as
\begin{equation}
\bb{x}_k =\bb{\kappa} \int_{t_{s,k}}^{t_{e,k}}\int_0^\tau C_p(\tau_1)d\tau_1d\tau + \bb{b}\int_{t_{s,k}}^{t_{e,k}}C_p(\tau)d\tau.
\label{eq:x_k}
\end{equation}
Putting $T$ time frames together, we can have the matrix format of equation (\ref{eq:x_k}) as 
\begin{equation}
\bb{x}(\bb{\theta}) = \bb{\theta}\bb{A}_{p}^{\mathsf{T}}, 
\end{equation}
where  $\bb{A}_p \in \mathbb{R}^{T \times 2}$ denotes the Patlak temporal matrix, with the $i$th row of $\bb{A}_p$ being $[ \int_{t_{s,i}}^{t_{e,i}}\int_0^\tau C_p(\tau_1)d\tau_1d\tau, \int_{t_{s,i}}^{t_{e,i}}C_p(\tau)d\tau]$. 
\medskip
\subsubsection{Proposed framework} 
Previously we have developed a direct Patlak reconstruction method based on the  linear kernel representation: $\bb{\theta} =  \bb{K}\bb{\delta}$ \cite{gong2017direct}, where $\bb{K}  \in \mathbb{R}^{N \times N} $ is the kernel matrix calculated based on the prior image and $\bb{\delta}  \in \mathbb{R}^{N\times 2}$ stands for the kernel-coefficient images.  The main idea is to represent the unknown parametric images by a linear combination of transformed features calculated from the prior information. Recently it was shown that instead of exploiting linear representation, nonlinear image representation using CNN can generate better results \cite{gong2018dip}. In this work, we proposed to represent the dynamic PET images generated based on the Patlak model by a CNN as
\begin{equation}
 \bb{\theta}\bb{A}_p ^{\mathsf{T}} = f(\bb{\alpha} | \bb{z}), 
\label{eq:cnn_repre}
\end{equation}
where $f:\mathbb{R}^N\rightarrow\mathbb{R}^{N\times T}$ represents the neural network, $\bb{\alpha}  \in \mathbb{R}^{S} $ are the unknown neural network parameters,  and $\bb{z} \in \mathbb{R}^{N}$ denotes the prior image from the same patient which was supplied as the network input. Note that for the network $f(\bb{\alpha} | \bb{z})$,  it can generate parametric images as the intermediate output and the final output will be dynamic PET images (more details explained in Sec.~\ref{network}). Based on (\ref{eq:systemmodel}), the dynamic PET system model can thus be rewritten as 
\begin{equation}
\bb{\bar{y}}(\bb{\alpha}) =  \boldsymbol{P}f(\bb{\alpha} | \bb{z})  + \bb{r}.
\end{equation} 
Through the CNN representation shown in  (\ref{eq:cnn_repre}), the task of reconstructing the unknown parametric image $\bb{\theta}$ was transferred to finding the network parameters $\hat{\bb{\alpha}}$ that maximized the likelihood function 
\begin{equation}
 L(\boldsymbol{y}|\boldsymbol{\alpha}) \propto \sum_{k=1}^T\sum_{i=1}^M (\bb{y}_k)_i \log (\bar{\bb{y}}(\bb{\alpha})_k)_i - (\bar{\bb{y}}(\bb{\alpha})_k)_i.
\label{orig_like}
\end{equation}
In  $L(\boldsymbol{y}|\boldsymbol{\alpha})$, the system matrix $\bb{P}$ is coupled with the CNN  $f(\bb{\alpha} | \bb{z})$, which is difficult to implement as $\bb{P}$ needs to be embedded in the network graph. In addition, the training speed will be slow as PET forward and backward projections are very time-consuming. The ADMM algorithm was employed to decouple $\bb{P}$ and $f(\bb{\alpha} | \bb{z})$. By introducing an auxiliary variable $\bb{v}\in \mathbb{R}^{N \times T}$, the original optimization in (\ref{orig_like}) can be transferred to
\begin{equation}
{\text{max}}_{\bb{\alpha},\bb{v}}\ L(\bb{y}|\bb{{v}}) ,\ \text{s.t.}\ \bb{{v}} = {f}({\bb{\alpha}} | \bb{z}).
\label{contrained}
\end{equation}
The constrained problem (\ref{contrained}) can be further transferred to an unconstrained optimization, solved through alternatively solving the following three subproblems:
\begin{align}
\bb{{v}}^{n+1} &= \argmaxA_{\bb{{v}}}  L(\bb{y}|\bb{{v}})  + Q(\bb{v})\label{eq:sub1} \\
\bb{\alpha}^{n+1} &= \argminA_{\bb{\alpha}} \Vert {f}({\bb{\alpha}} | \bb{z})- (\bb{v}^{n+1} + \bb{\mu}^n)\Vert^2,  \label{eq:sub2} \\
\bb{\mu}^{n+1} &= \bb{\mu}^n + \bb{v}^{n+1} - f({\bb{\alpha}^{n+1}} | \bb{z}),  \label{eq:sub3}
\end{align}
where
\begin{equation}
 Q(\bb{v}) = - \frac{\rho}{2} \Vert\bb{{v}} -{f}({\bb{\alpha}^n} | \bb{z}) + \bb{\mu}^n\Vert^2. 
\end{equation}
Note that subproblem (\ref{eq:sub1}) is a frame-by-frame penalized image reconstruction problem, which can be solved using existing PET static reconstruction algorithms. Optimization transfer \cite{lange2000optimization} was chosen to solve it in our work. The surrogate function for $L(\bb{y}|\bb{{v}})$ regarding frame $t$ and voxel $j$ is 
\begin{equation}
\varphi(v_{jt}|\bb{v}^{n}) = p_j (\hat{v}^{n+1}_{jt,\textrm{EM}} \log v_{jt} - v_{jt}), 
\label{eq:surrogate_lyx}
\end{equation}
where $p_{j} = \sum_{i=1}^{M}P_{ij}$ and $\hat{v}^{n+1}_{jt,\textrm{EM}}$ was calculated by 
\begin{equation}
\hat{v}^{n+1}_{jt,\textrm{EM}} = \frac{v^n_{jt}}{p_{j}} \sum_{i=1}^{M}P_{ij}\frac{y_{it}}{ [\bb{P}\bb{v}^n]_{it} + r_{it}}.
\label{admm_scale}
\end{equation}
The final iterative update equation for Subproblem (\ref{eq:sub1}) can thus be obtained by setting the first gradient of $\varphi(v_{tj}|\bb{v}^{n})+ Q(v_{jt})$ to 0. Subproblem (\ref{eq:sub2}) is a network training problem based on a L2-norm loss. In our work, the L-BFGS algorithm was employed for the network training problem (running 20 epochs per loop) due to its monotonic property.
\begin{figure}[t]
\centering
{\includegraphics[trim=7.6cm 2cm 7.2cm 3.0cm, clip,width=3in]{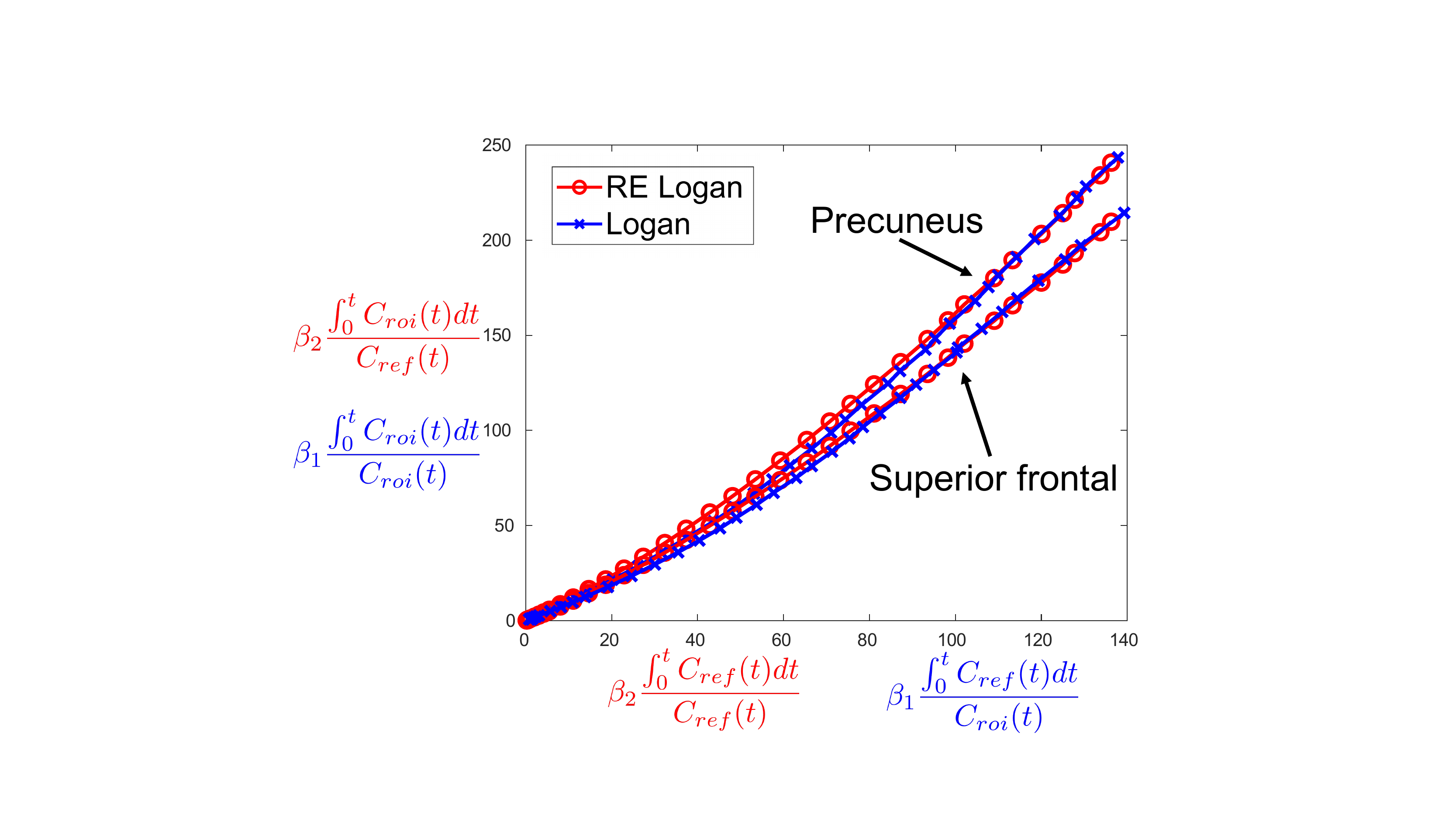}}
\caption{\small{The plots comparing Logan (blue curves) and RE Logan (red curves) models based on the precuneus and superior frontal cortices. The $x$ and $y$ axes were scaled to match the curves from the two models to better observe the slopes.} } 
\label{fig:validation_relogan}
\end{figure}

\begin{figure*}[t]
\centering
\subfloat{\includegraphics[trim=5.5cm 5.5cm 5.4cm 7.2cm, clip, width=5.6in]{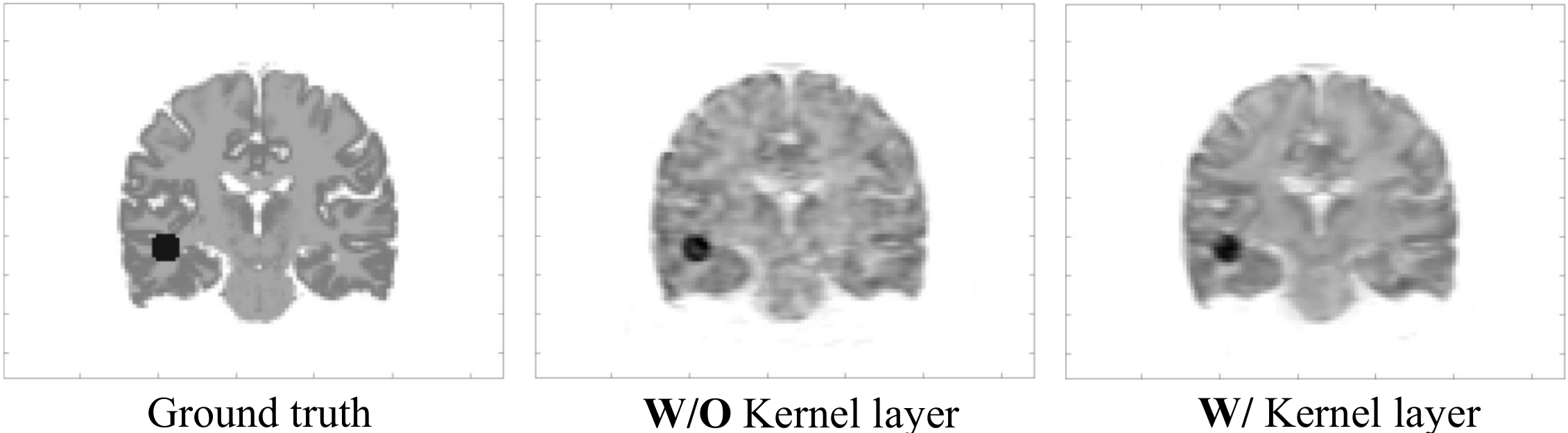}}
\caption{\small{Comparisons of the network output w/o and w/ the kernel layer. The left column is the ground-truth Patlak-slope image. For both scenarios, the network was trained with the L-BFGS algorithm running 1000 epochs. } }
\label{fig:kernel_layer}
\vspace{-0.3cm}
\end{figure*}

\subsection{ Proposed framework for the RE Logan model}
\subsubsection{RE Logan model}
For reversible tracers, the Logan  model \cite{logan1996distribution}  is widely used. According to the Logan model, after reaching a steady time $t_1^{\ast}$, the tracer concentration image $\bb{c}(t;\bb{\theta})$ can be written as 
\begin{equation}
\frac{\int_{0}^{t}\bb{c}(\tau;\bb{\theta})d\tau}{\bb{c}(t;\bb{\theta})} = \bb{DV} \frac{\int_{0}^{t}C_{ref}(\tau)d
\tau}{\bb{c}(t;\bb{\theta})} + \bb{q}, t \geq t_1^{\ast}
\label{logan_model}
\end{equation}
where the division operation is element-wise, $\bb{DV}\in \mathbb{R}^{N}$ denotes the distribution volume (DV) image,  $\bb{q}\in \mathbb{R}^{N}$ is the intercept image,  $C_{ref}(t)$ is the tracer concentration of the reference region, and $\bb{\theta} = [\bb{DV}, \bb{q}]$. Different from the Patlak model, directly embedding the Logan model into the direct reconstruction framework is difficult as $\bb{c}(t; \bb{\theta})$ is coupled across different time frames due to the integration process. Here we used the relative equilibrium version of the Logan model, the RE Logan model \cite{zhou2009consistent}, as it can be easily embedded into the direct reconstruction framework. The RE Logan model is based on the assumption that there exists $t_2^{\star}$ such that the tracer concentrations in all tissue compartments reach equilibrium relative to plasma input for $t\geq t_2^{\ast}$. Based on the RE Logan model,
\begin{equation}
\frac{\int_{0}^{t}\bb{c}(\tau;\bb{\theta})d\tau}{C_{ref}(t)} = \bb{DV} \frac{\int_{0}^{t}C_{ref}(\tau)d\tau}{C_{ref}(t)} + \bb{q}, t \geq t_2^{\ast}.
\label{eq:relogan}
\end{equation}
Based on~(\ref{eq:relogan}), we can further get
\begin{equation}
 \sum_{i= 1}^k \bb{x}_i=  \bb{DV}  \int_{0}^{t_{e,k}}C_{ref}(\tau)d\tau+ \bb{q}C_{ref}(t_{e,k}).
 \label{eq:relogan_tac}
\end{equation}
\subsubsection{Proposed framework}
One way to embed~(\ref{eq:relogan_tac}) into the direct reconstruction framework is through combining sinograms from  frame 1 to $k$  for the corresponding  $ \sum_{i= 1}^k \bb{x}_i$ image. However, this will violate the i.i.d. assumption of the sinogram events. To solve this issue, we first proposed to  combine the frames from $t=0$ to $t = t_2^{\ast}$ as the new $\bb{x}_1$. This new framing strategy is further explained in Fig.~\ref{fig:data_bining}. Based on this new framing, we proposed a direct reconstruction framework for the RE Logan model based on the following constrained optimization 
\begin{equation}
\argmaxD   \ L(\boldsymbol{y}|\bb{v})  \quad \mbox{s.t.} \  \bb{v} \bb{B}=   \bb{\theta} \bb{A}_r^{\mathsf{T}} ,
\label{eq:direct_logan}
\end{equation}
where 
\begin{equation*}
\bb{B} = 
\begin{pmatrix}
1 & 1  & \cdots & 1 \\
0 & 1  & \cdots & 1 \\
\vdots  & \vdots  & \ddots & \vdots  \\
0 & 0  & \cdots & 1
\end{pmatrix}
\end{equation*}
is a $T \times T$ matrix to  combine different time frames, and $\bb{A}_r \in \mathbb{R}^{T \times 2}$ denotes the RE Logan temporal matrix, with the $i$th row of $\bb{A}_r$ being $[ \int_{0}^{t_{e,i}}C_{ref}(\tau)d\tau, C_{ref}(t_{e,i})]$. 

In this work, to leverage the high-quality prior image, similar to the Patlak model,  we proposed to represent the dynamic PET images generated through the RE Logan model by the output of a CNN as $ \bb{\theta} \bb{A}_r^{\mathsf{T}} = {f}({\bb{\alpha}}|\bb{z})$. The objective function of the proposed direct RE Logan reconstruction in (\ref{eq:direct_logan}) can  be written as 
\begin{equation}
\argmaxD L(\boldsymbol{y}|\bb{v}) \quad   \mbox{s.t.} \:  \bb{v} \bb{B} =   {f}({\bb{\alpha}}|\bb{z}).
 \label{eq:relog_obj}
\end{equation}
Based on the ADMM algorithm, (\ref{eq:relog_obj}) can be decomposed into the following subproblems as:
\begin{align}
\bb{v}^{n+1} &= \argmaxA_{\bb{v}}  L(\bb{y}|\bb{v}) 
+ Q(\bb{v}), \label{eq:logansub_1}\\
{\bb{\alpha}}^{n+1} &= \argminA_{{\bb{\alpha}}} \Vert   {f}({\bb{\alpha}}|\bb{z})-\bb{v}^{n+1}\bb{B} - \bb{\mu}^n\Vert^2, \label{eq:logansub_2} \\
\bb{\mu}^{n+1} &= \bb{\mu}^n + \bb{v}^{n+1}\bb{B} -   {f}({\bb{\alpha}}^{n+1}|\bb{z}), 
\label{eq:logansub_3}
\end{align}
where 
\begin{equation}
Q(\bb{v}) = - \frac{\rho}{2} \Vert\bb{v}\bb{B} -    f(\bb{\alpha}^n|\bb{z} )+ \bb{\mu}^n\Vert^2.
\end{equation}
For Subproblem (\ref{eq:logansub_1}), due to the time-domain coupling  ($\bb{v}\bb{B}$ part), frame-by-frame reconstruction cannot be conducted directly. The optimization transfer algorithm was used to transfer it to pixel-by-pixel and frame-by-frame reconstruction. For $Q(\bb{v})$, the surrogate function chosen at iteration $n$ for voxel $j$ and frame $t$ is
 \begin{align}
  \Psi(v_{jt}| \bb{v}^n) &=  - \frac{\rho}{2}\sum_{i=1}^T\frac{b_{it}v^n_{jt}}{[\bb{v}^n_{j\cdot}\bb{B}]_i}\Bigg(\Bigg.\frac{[\bb{v}^n_{j\cdot}\bb{B}]_i v_{jt}}{v^n_{jt}}\nonumber \\
 &- [ {f}({\bb{\alpha}}^{n}|\bb{z})]_{jt} + {\mu}^n_{jt} \Bigg.\Bigg)^2.
 \end{align}
The final iterative update equation for Subproblem (\ref{eq:logansub_1}) can thus be obtained by setting the first gradient of $\varphi(v_{jt}|\bb{v}^{n})+   \Psi(v_{jt}| \bb{v}^n)$ to 0, where $\varphi(v_{tj}|\bb{v}^{n})$ is given in (\ref{eq:surrogate_lyx}). Subproblem (\ref{eq:logansub_2}) is a network training problem based on a L2-norm loss similar to (\ref{eq:sub2}), which was also solved with the L-BFGS algorithm running 20 epochs per loop.  

\subsection{ Network structure}\label{network}
The schematic plot of the network structure ${f}({\bb{\alpha}}|\bb{z})$ is presented in Fig.~\ref{fig:network_structure}.  It consists of a 3D U-net structure with a proposed kernel layer embedded to generate the parametric images, and a kinetic model-based convolution layer to output the dynamic PET images. The input to the network $\bb{z}$ is the T1-weighted MR image from the same patient. More detailed explanations about the network design are as follows.

For the operation of $\bb{\theta} \bb{A}^{\mathsf{T}}$,  $\bb{A} \in \mathbb{R}^{2\times T}$,  it can be interpreted as a convolution operation with a  $1\times1 \times 2 \times T$ convolution kernel.  For the Patlak model and the RE Logan model,  $\bb{A}$  is  $\bb{A}_p$  and  $\bb{A}_r$, respectively. Based on this observation,  the linear kinetic models can be implemented as convolution layers with pre-calculated weights in the network graph. These kinetic-model layers need to be deployed as the last layer before the network output so that the network ${f}({\bb{\alpha}}|\bb{z})$ can generate the parametric images as the intermediate output.

The 3D Unet structure \cite{cciccek20163d} was adopted as the backbone of ${f}({\bb{\alpha}}|\bb{z})$ in this work. We further designed a kernel layer, inspired by  the kernel method \cite{guobao15}, to better leverage the high-resolution prior image $\bb{z}$ widely available in PET imaging. The kernel method has been successfully applied to various prior image-guided PET image reconstruction problems, where the unknown image $\bb{x}$ is represented as $\bb{x} = \bb{K}\bb{\delta}$. If the kernel matrix $\bb{K}$ is constructed by the radial basis function,  the operation of $\bb{K}\bb{\delta}$ is equivalent to a nonlocal denoising operation. Inspired by this, we proposed to construct a kernel layer to perform nonlocal feature denoising as
\begin{equation}
\bb{x}_{\text{out}}= \bb{K} \bb{x}_{\text{in}},
\end{equation} 
where $\bb{x}_{\text{in}} \in \mathbb{R}^{ N \times C}$ is the kernel-layer input with $C$ being the feature size, $\bb{x}_{\text{out}} \in \mathbb{R}^{ N \times C} $ is the kernel-layer output, and $\bb{K} \in \mathbb{R}^{ N \times N} $ is the kernel matrix which contains the similarity coefficients constructed from the prior structural image $\bb{z}$. Note that the same prior image $\bb{z}$ was also supplied as the network input. The $(i,j)$th element of the kernel matrix $\boldsymbol{K}$ was calculated as
\begin{equation}
k_{ij} = \exp\left(-\frac{||\boldsymbol{f}_i-\boldsymbol{f}_j||^2}{2N_f\sigma^2}\right),
\end{equation}
where $\bb{f}_i \in \mathbb{R}^{N_f}$ and $\bb{f}_j \in \mathbb{R}^{N_f}$ are the feature vectors of voxel $i$ and voxel $j$ from the prior image $\bb{z}$, respectively, $\sigma^2$ is the variance of $\bb{z}$ and $N_f$ is the number of voxels in a feature vector. A 3$\times$3$\times$3 local patch was extracted for each voxel to construct the feature vector ($N_f = 27$). Instead of saving all the $k_{ij}$ elements, the kernel matrix was constructed using a $K$-Nearest-Neighbor ($K$NN) search in a 7$\times$7$\times$7 search window with 50 elements saved to make $\bb{K}$ sparse. $\bb{K}^{\mathsf{T}}$ was also calculated to enable back-propagation of the kernel layer. One concern of utilizing MR prior is the potential mismatch regions between PET and MR images.  Thus, instead of putting the kernel layer at the end of the network, we push it inside the network several more blocks to help better recover potential mismatch regions.
\begin{figure*}[t]
\vspace{-0.5cm}
\centering
\subfloat{\includegraphics[trim=0cm 0cm 0cm 0.0cm, clip, width=5in]{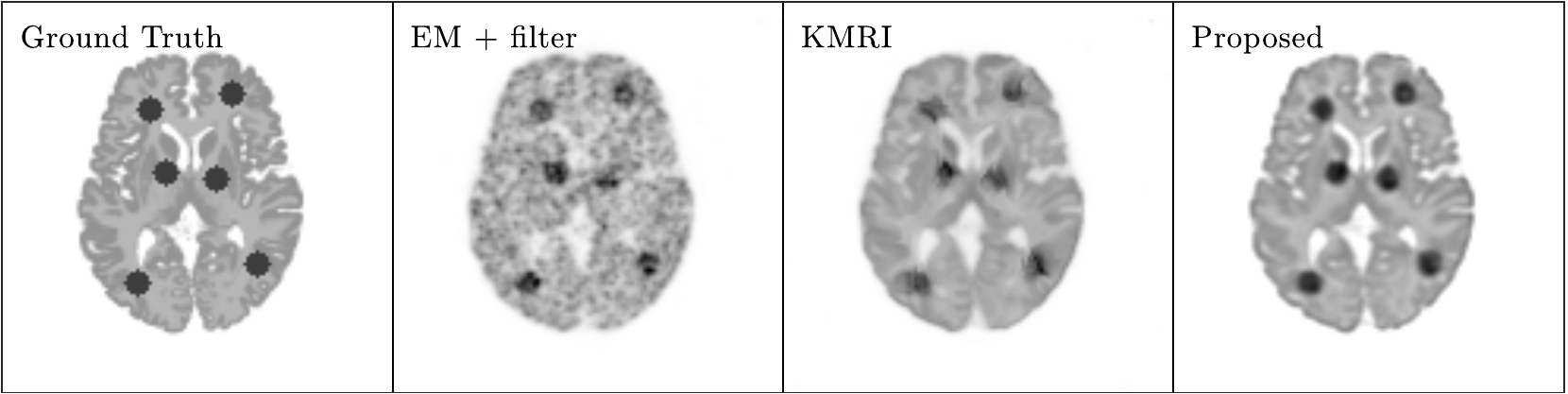}}\\
\subfloat{\includegraphics[trim=0cm 0cm 0cm 0.0cm, clip, width=5in]{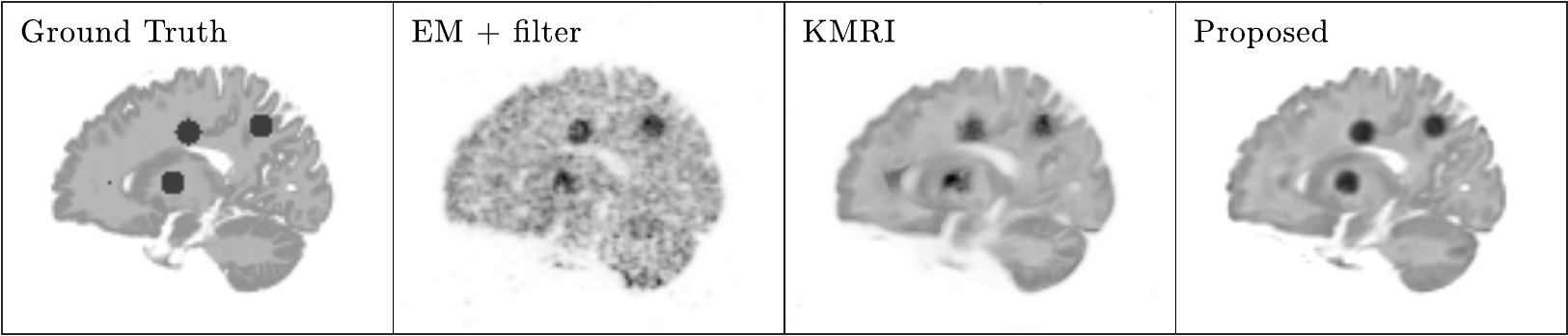}}\\
\subfloat{\includegraphics[trim=0cm 0cm 0cm 0.0cm, clip, width=5in]{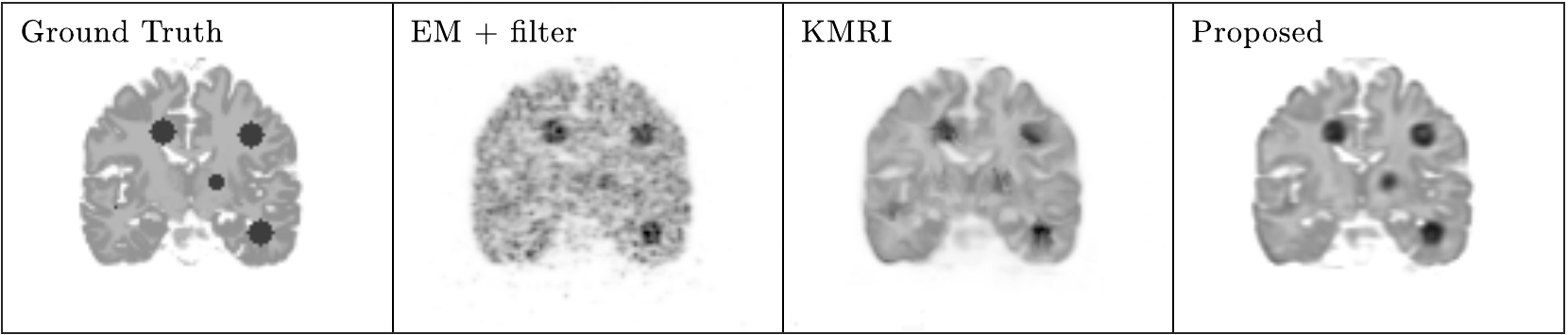}}\\
\caption{\small{Different views of the reconstructed Patlak slope image using different methods for the simulation study. The first column is the ground-truth image.} }
\vspace{-0.4cm}
\label{fig:simu_img_appear}
\end{figure*}
\begin{figure*}[t]
\centering
\subfloat{\includegraphics[trim=0cm 0cm -0.4cm 0cm, clip, width=2.5in]{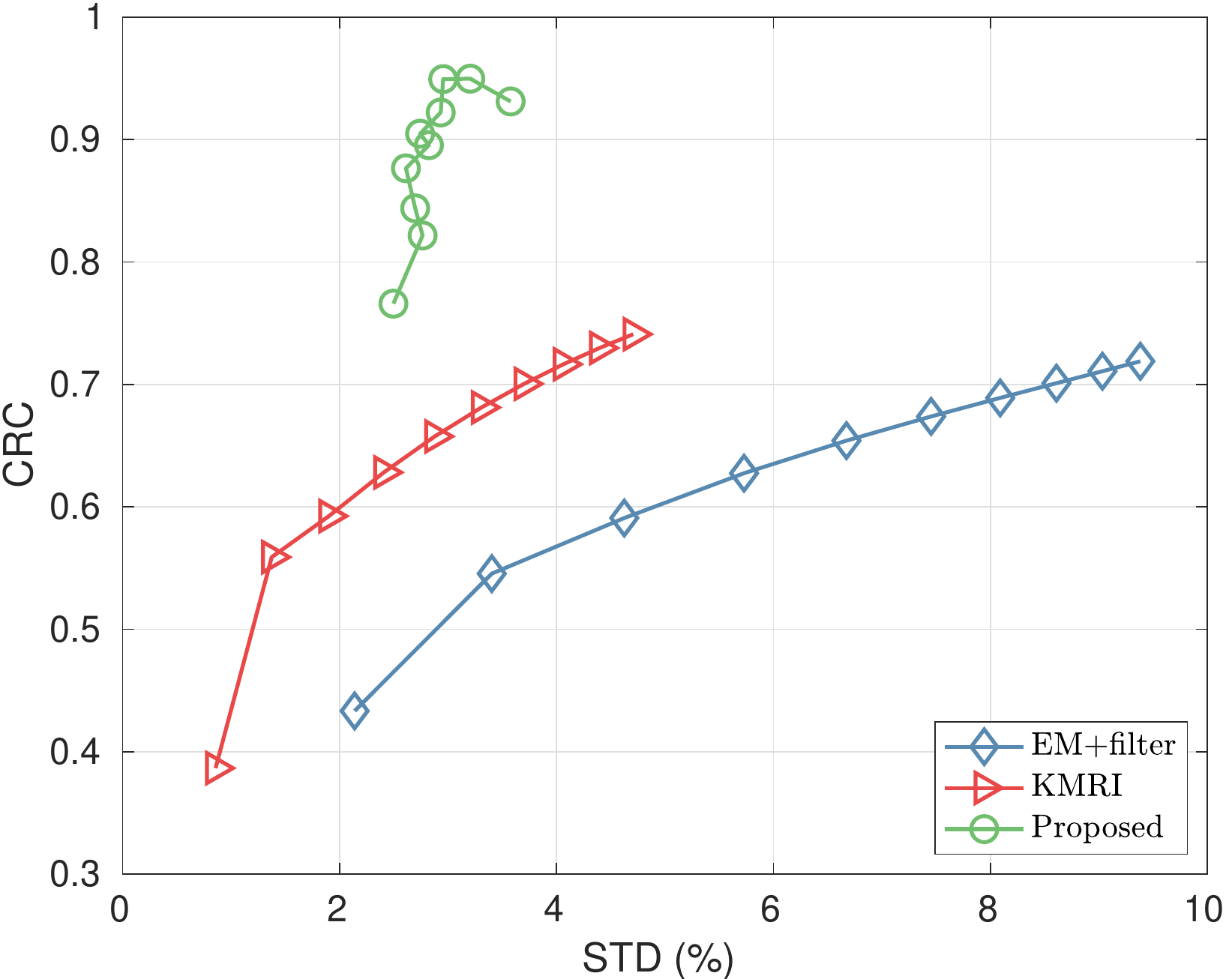}}
\subfloat{\includegraphics[trim=-0.4cm 0cm 0cm 0cm, clip, width=2.5in]{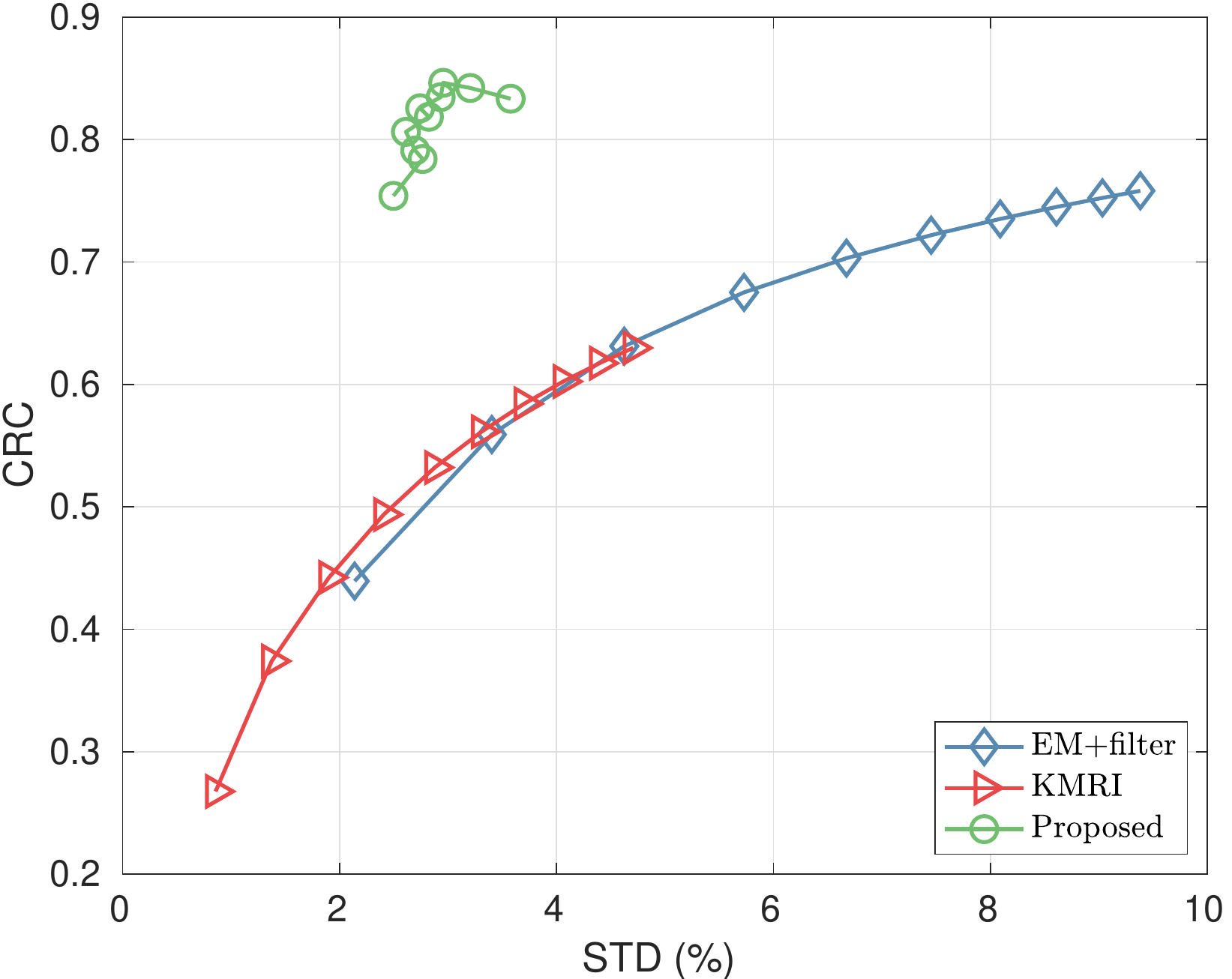}}
\caption{\small{CRC vs. STD for (left) the gray matter ROIs and (right) the artificially inserted tumor regions at different iteration numbers. } }
\label{fig:simu_quantitative}
\vspace{-0.3cm}
\end{figure*}
\begin{figure*}[t]
\vspace{-0.5cm}
\centering
\subfloat{\includegraphics[trim=1.8cm 0cm 1.8cm 0cm, clip, width=1.32in]{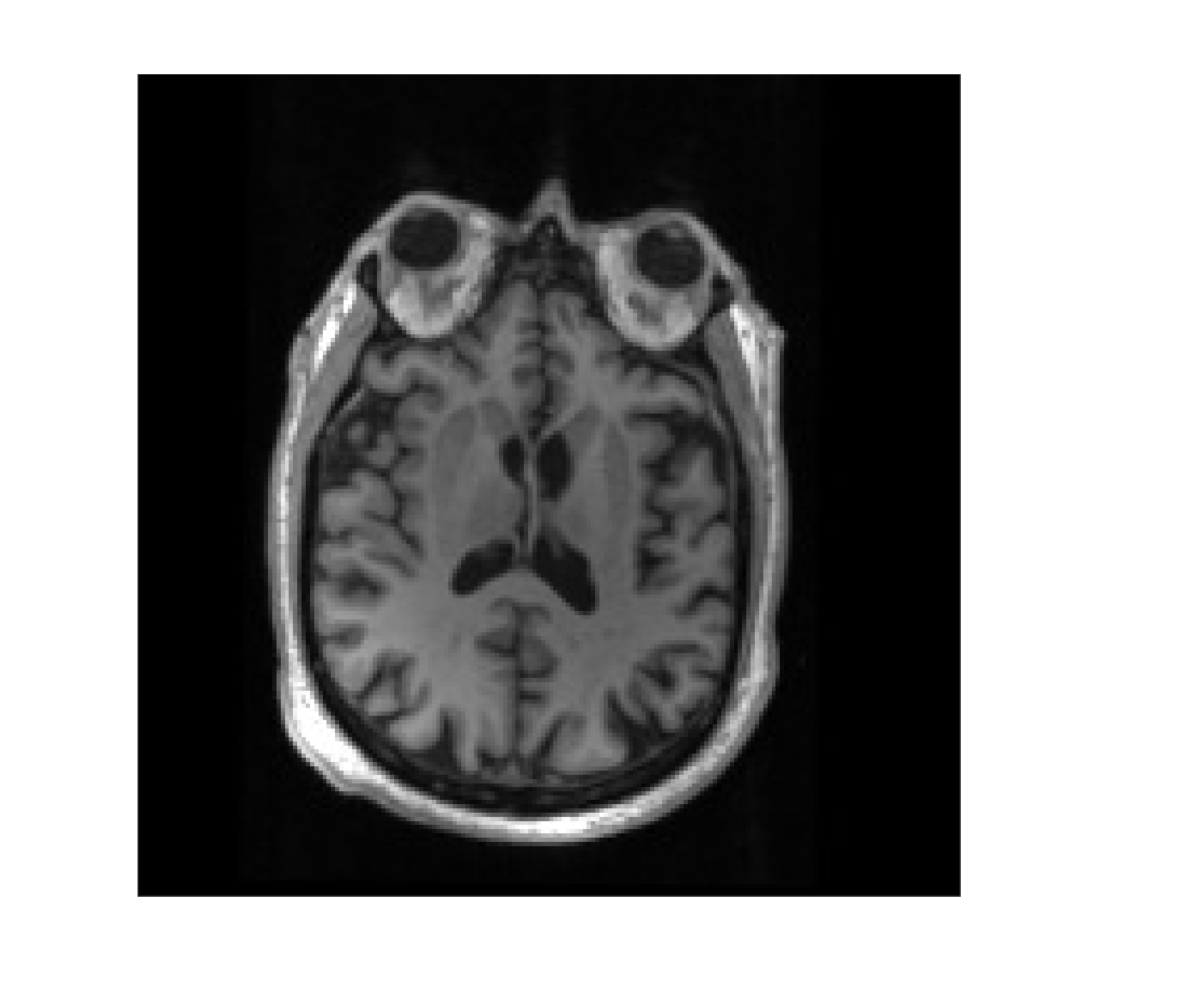}}\hspace{-0.3cm}
\subfloat{\includegraphics[trim=2.6cm 0cm 4.2cm 0.0cm, clip, width=3.5in]{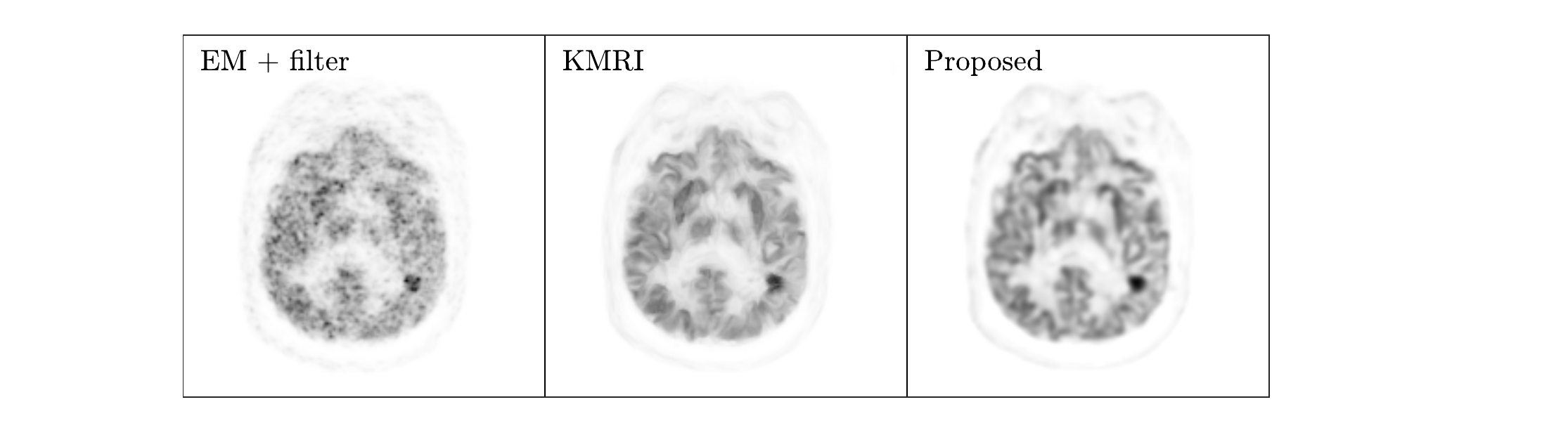}}\vspace{-0.7cm}\\
\subfloat{\includegraphics[trim=1.8cm 0cm 1.8cm 0cm, clip, width=1.32in]{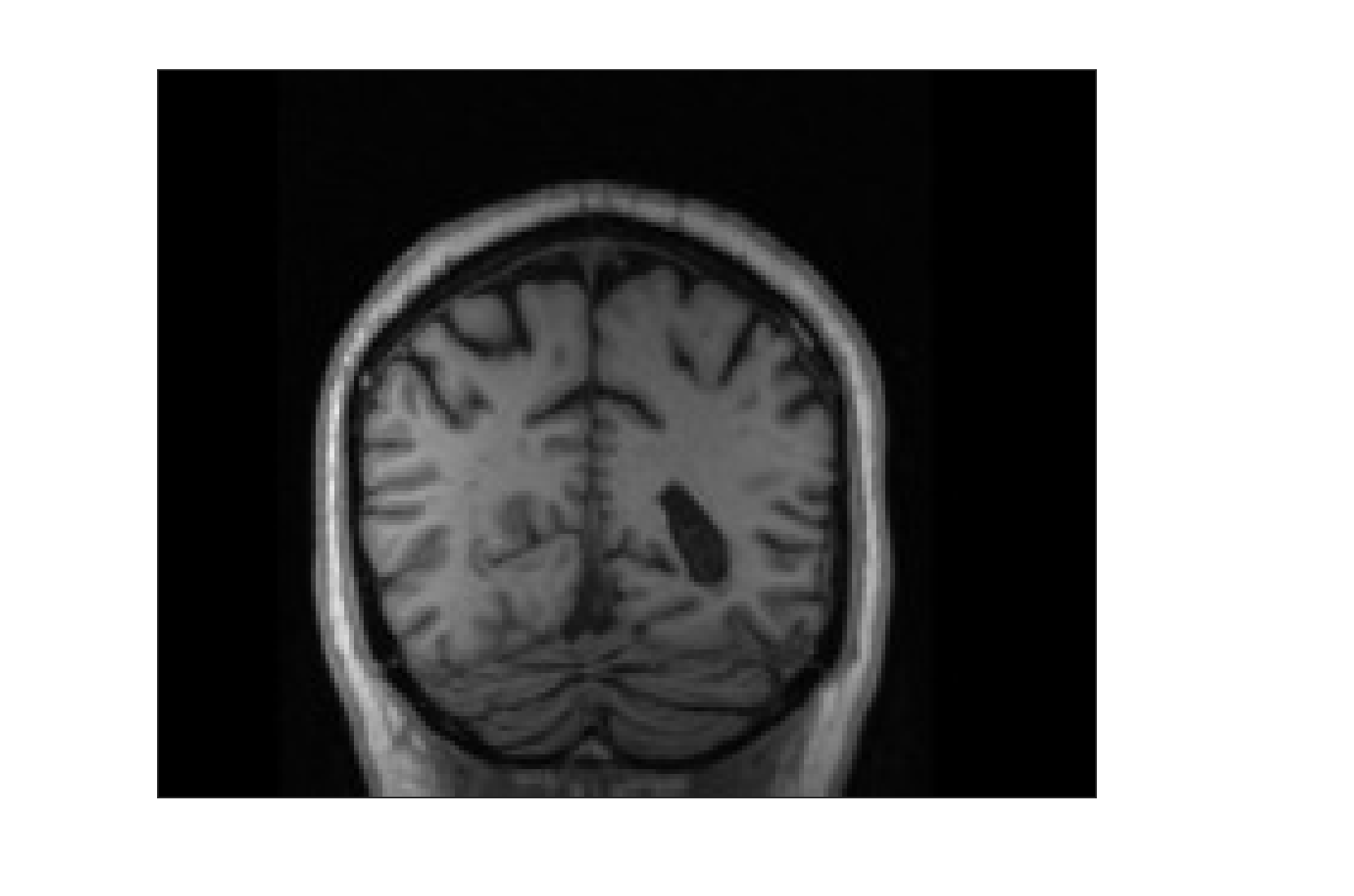}}\hspace{-0.3cm}
\subfloat{\includegraphics[trim=2.6cm 0cm 4.2cm 0.0cm, clip, width=3.5in]{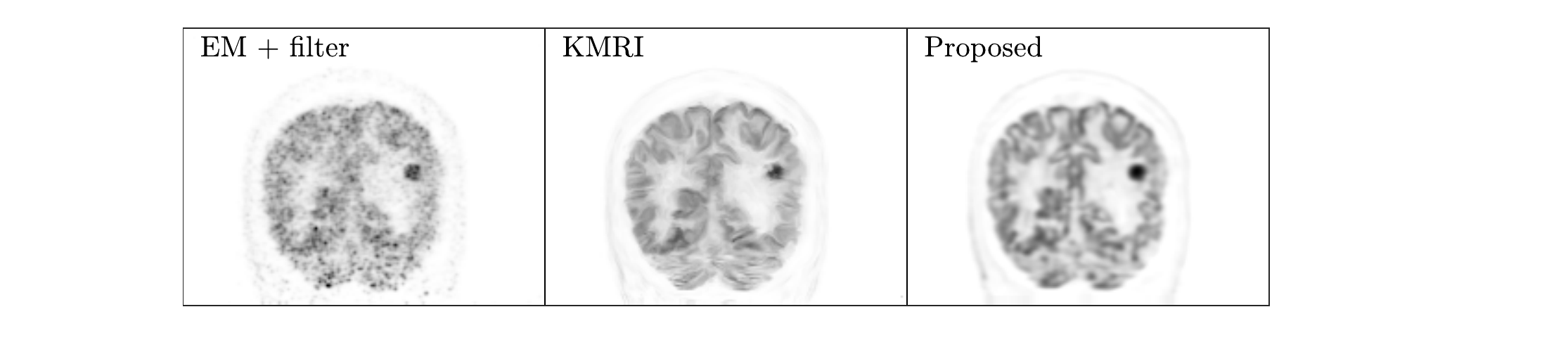}}\vspace{-0.63cm}\\
\subfloat{\includegraphics[trim=1.8cm 0cm 1.8cm 0cm, clip, width=1.32in]{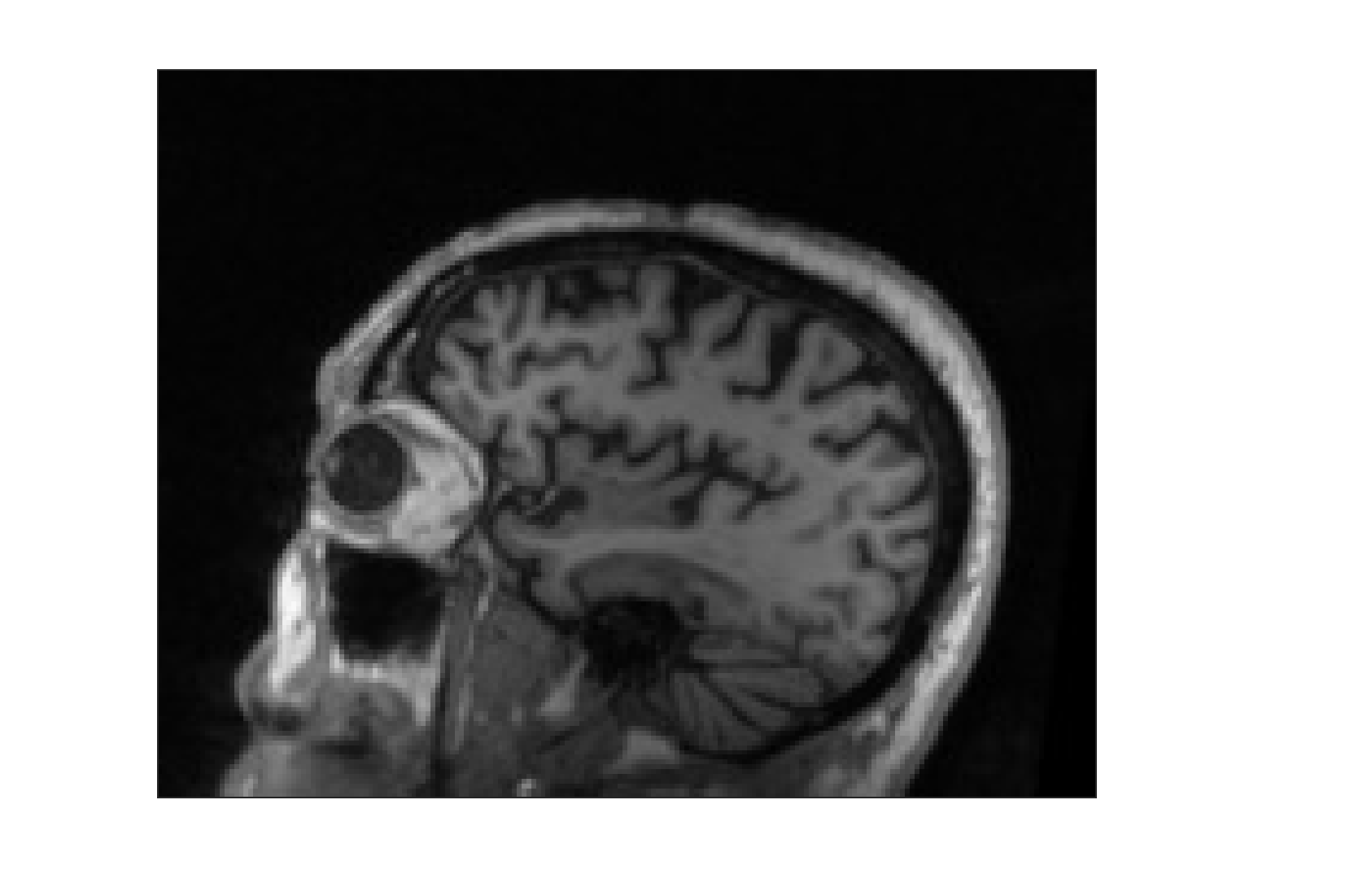}}\hspace{-0.3cm}
\subfloat{\includegraphics[trim=2.6cm 0cm 4.2cm 0.0cm, clip, width=3.5in]{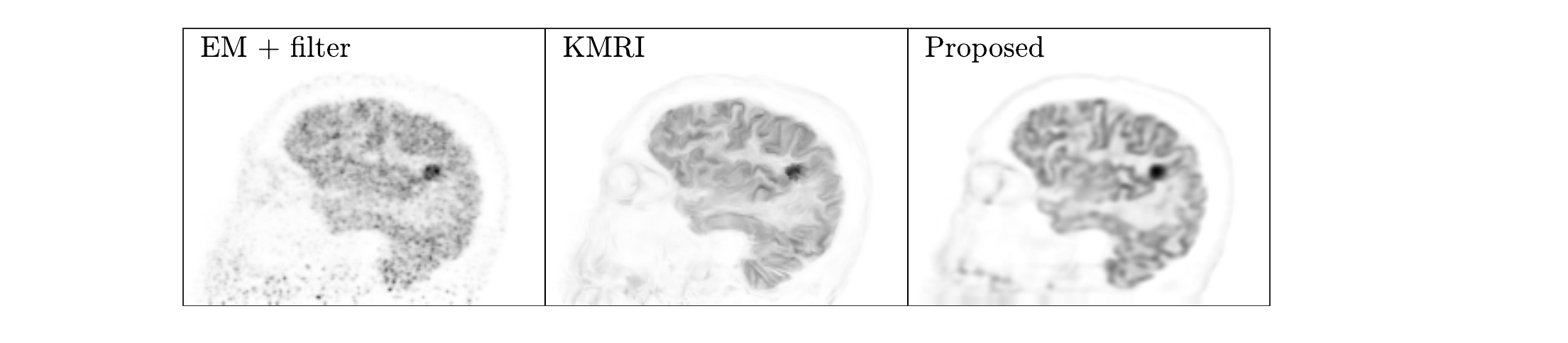}}\vspace{-0.63cm}\\
\subfloat{\includegraphics[trim=1.8cm 0cm 1.8cm 0cm, clip, width=1.32in]{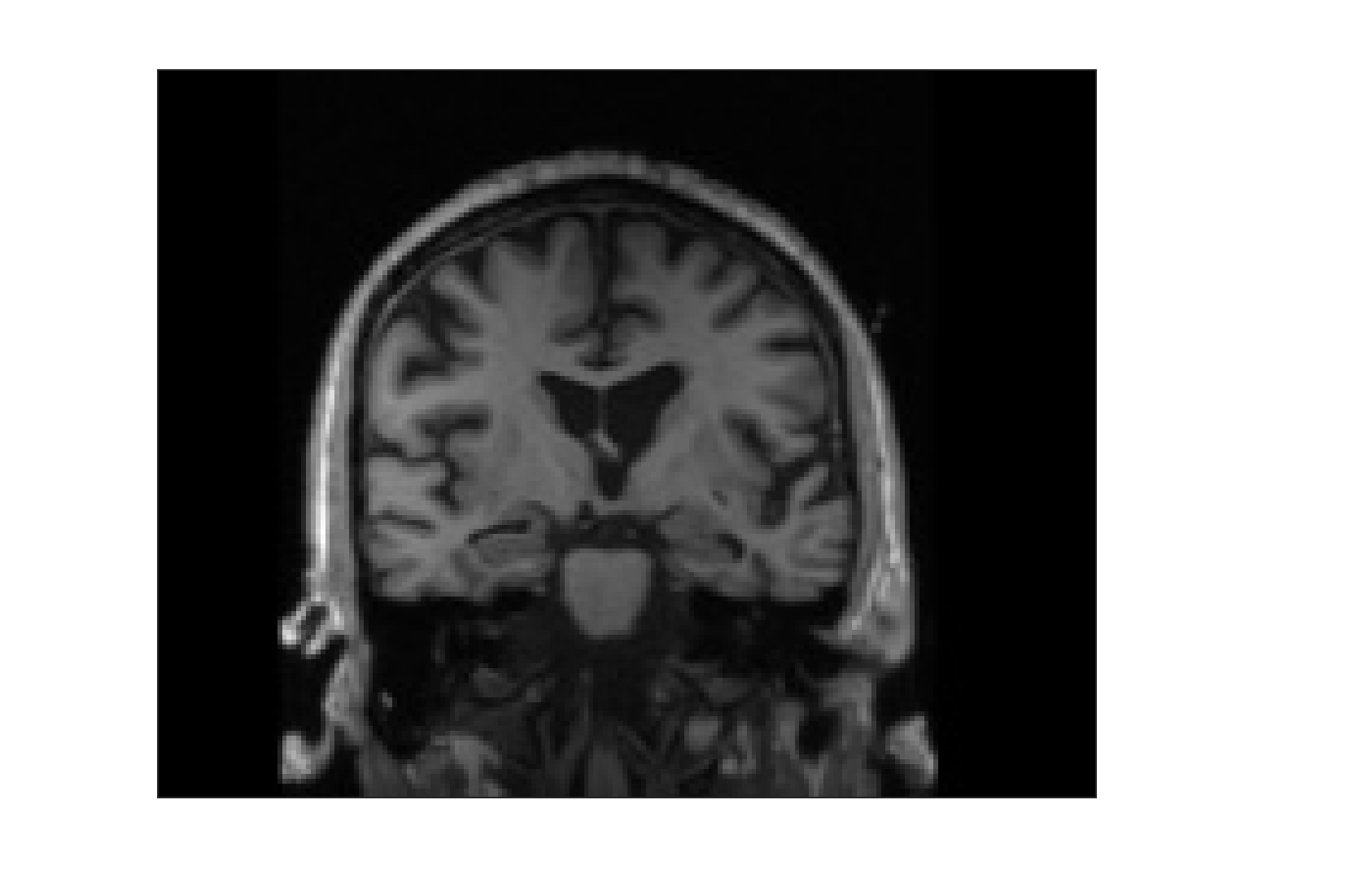}}\hspace{-0.3cm}
\subfloat{\includegraphics[trim=2.6cm 0cm 4.2cm 0.0cm, clip, width=3.5in]{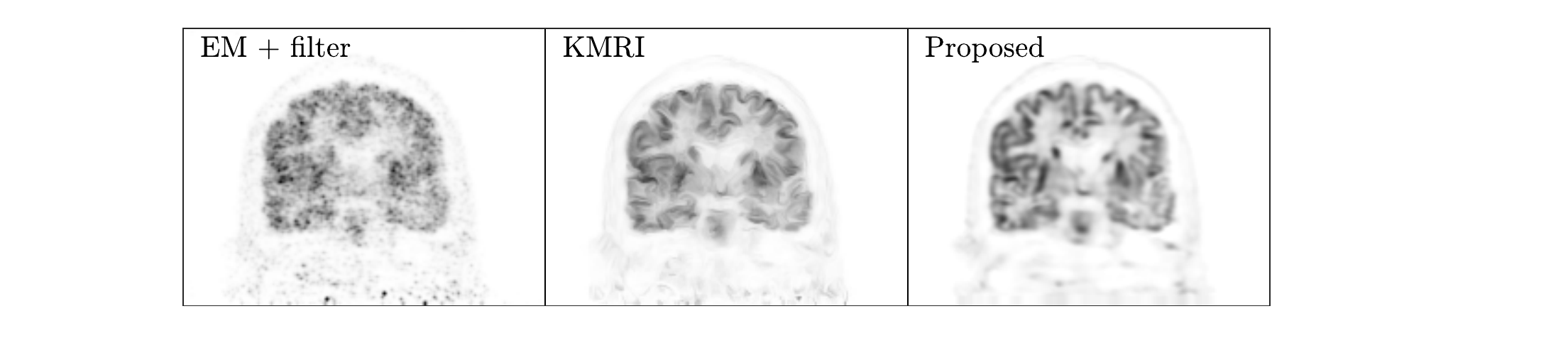}} \\
\caption{\small{Different views of the reconstructed Patlak slope image using different methods. The first column shows the corresponding T1-weighted MR prior image.} }
\vspace{-0.4cm}
\label{fig:real_img_appear}
\end{figure*}

\begin{figure*}[h]
\centering
\subfloat{\includegraphics[trim=0cm 0.cm -0.4cm 0cm, clip, width=2.4in]{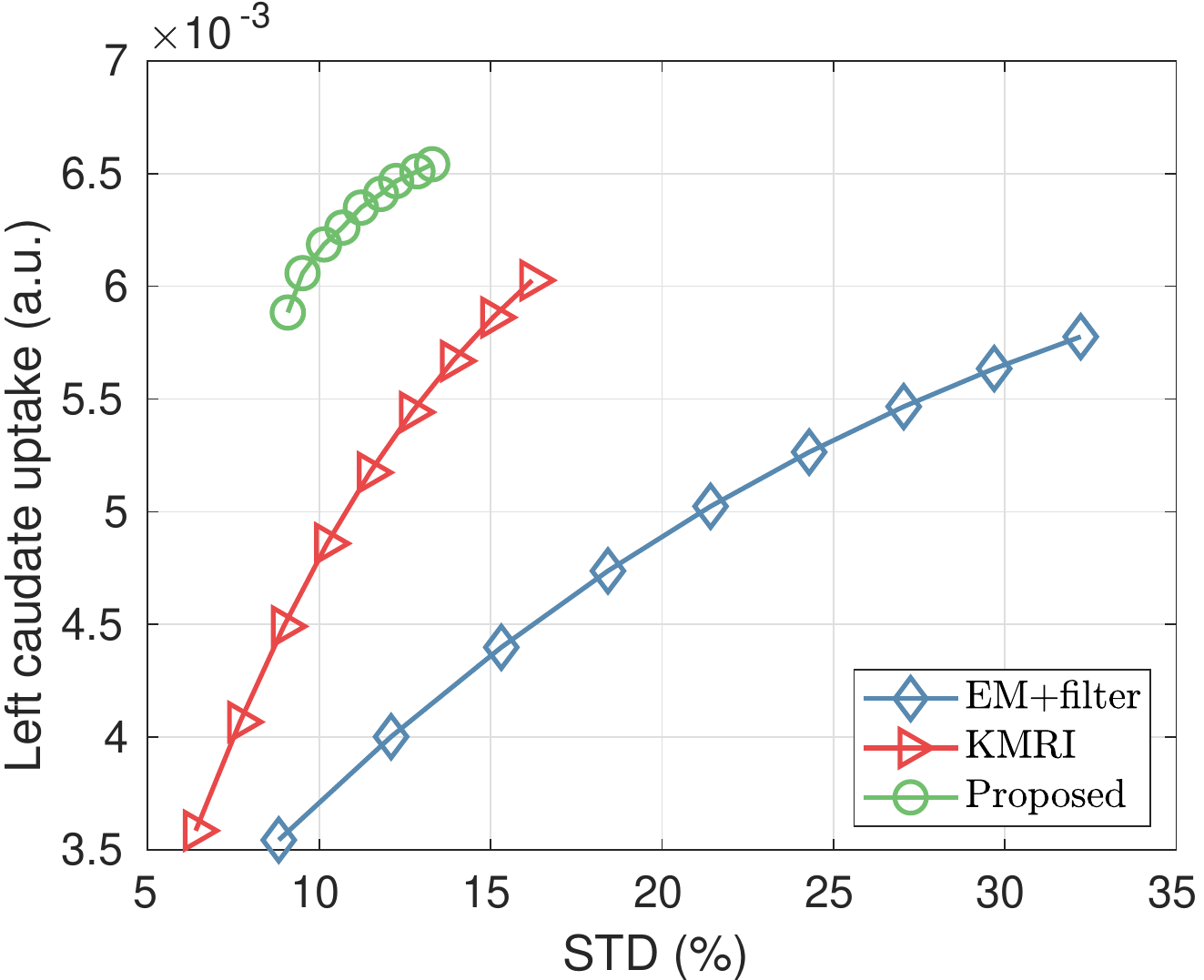}}
\subfloat{\includegraphics[trim=0cm 0.cm -0.40cm 0cm, clip, width=2.4in]{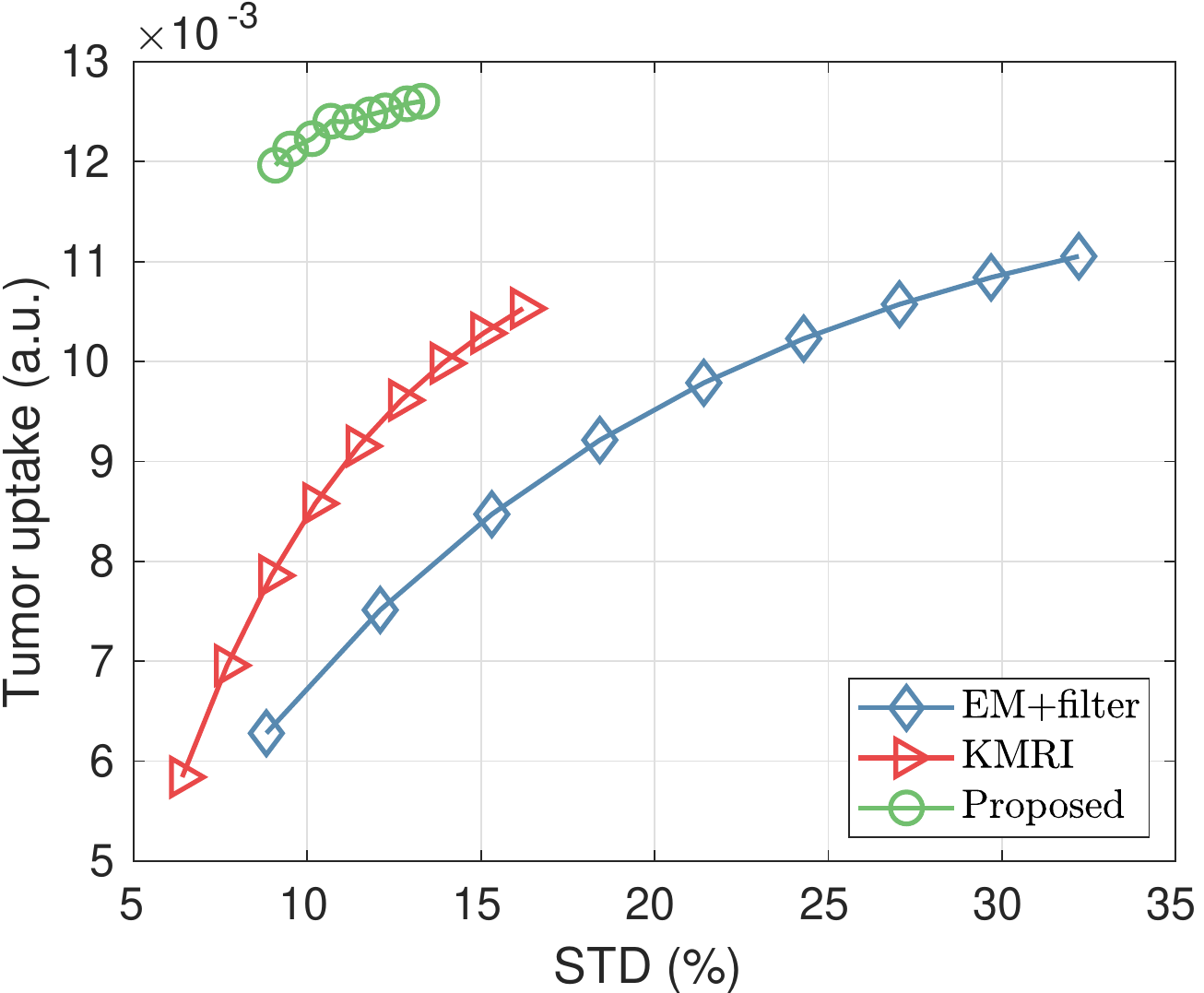}}
\caption{\small{Regional uptake vs. STD for (left) the left caudate ROI and (right) the artificially inserted tumor region at different iteration numbers. } }
\label{real_quantitative_patlak}
\vspace{-0.6cm}
\end{figure*}
The nonlocal operation in the proposed kernel layer has several differences compared to that in Wang {\it{et al}}'s work of nonlocal neural networks \cite{wang2018non}. Firstly, in  \cite{wang2018non}, the similarity was calculated from the feature vectors extracted from the previous layer. In the proposed kernel layer, the similarity was calculated based on the fixed prior image, which is widely available in PET imaging and has higher resolution and SNR than the extracted features. Secondly, the focus of \cite{wang2018non} is on image classification and the nonlocal operation was located close to the final network output, where the spatial size is much smaller than the original image. For denoising applications, the spatial size of the nonlocal operation should be similar to the original image in order to be effective. However,  for large spatial size, accurately learning the large-size embedding weights proposed in  \cite{wang2018non} is difficult due to training-data and GPU memory limits. In this work, the similarity calculation was based on the radial basis function. It did not involve training parameters and can be pre-calculated, which is especially suitable for unsupervised learning frameworks (no training data) and 3D denoising applications (large spatial size).

\subsection{Reference methods}

For the Patlak model, the direct reconstruction based on the nested EM algorithm with Gaussian post-filtering\cite{wang2010acceleration} was adopted as the baseline method, denoted as $\text{EM}+\text{filter}$. Additionally, the kernel method-based direct reconstruction was also utilized for comparison \cite{gong2017direct}, denoted as KMRI, where the kernel matrix was calculated the same as in the kernel layer. For the RE Logan model,  the direct reconstruction based on the objective function in (\ref{eq:direct_logan}) was adopted as the baseline method, denoted as $\text{Direct}+\text{filter}$. Based on the ADMM algorithm, the subproblems involved are similar to the proposed method by replacing the neural network representation ${f}({\bb{\alpha}}|\bb{z})$ with the parametric image itself. The kernel method was also developed for the direct RE Logan model as a reference method, denoted as KMRI. The subproblems involved are also similar to the proposed method by replacing the neural network representation ${f}({\bb{\alpha}}|\bb{z})$ with the kernel representation. The proposed network was implemented based on TensorFlow 1.16 on GPU V100. 

\section{Experiment}
\subsection{Simulation study for the Patlak model}\label{sec:simulation}
A 3D brain phantom from the Brainweb \cite{cocosco1997brainweb} was used in the simulation study based on the Siemens mCT scanner \cite{jakoby2011physical}. The system matrix $\bb{P}$ was computed using the multi-ray tracing method \cite{zhou2011fast}. The time activity curves of the gray matter and white matter were generated \txtc{mimicking an FDG scan} using the same set-up as in \cite{gong2017direct}. Twelve hot spheres of diameter 16 mm, not visible in the MR image, were inserted into the PET image as tumor regions to simulate mismatches between the MR and PET images. The dynamic PET scan was divided into 24 time frames: 4$\times$20 s, 4$\times$40s, 4$\times$60 s, 4$\times$180 s, and 8$\times$300 s. Noise-free sinogram data were generated by forward-projecting the ground-truth images using the system matrix and the attenuation map. Uniform random events were simulated and accounted for 30 percent of the noise free data in all time frames. Poisson noise was then introduced to the noise-free data by setting the total count level to be equivalent to an 1-hour 18F-FDG scan with 5 mCi injection. The reconstructed image has a matrix size of 125 $\times$ 125 $\times$ 105 and a voxel size of 2 $\times$ 2 $\times$ 2 $\mbox{mm}^3$. For the direct Patlak reconstruction, only the last 5 frames for a total duration of 25 minutes were used ($ t^{\ast} = 35 $ min). The contrast recovery coefficient (CRC) and the standard deviation (STD) based on 20 noise realizations were calculated the same way as in \cite{gong2017direct} for the gray matter and the tumor ROIs to perform quantitative comparisons.
\begin{figure*}[t]
\centering
{\includegraphics[trim=4.6cm 1cm 3.2cm 1.2cm, clip, width=6.6in]{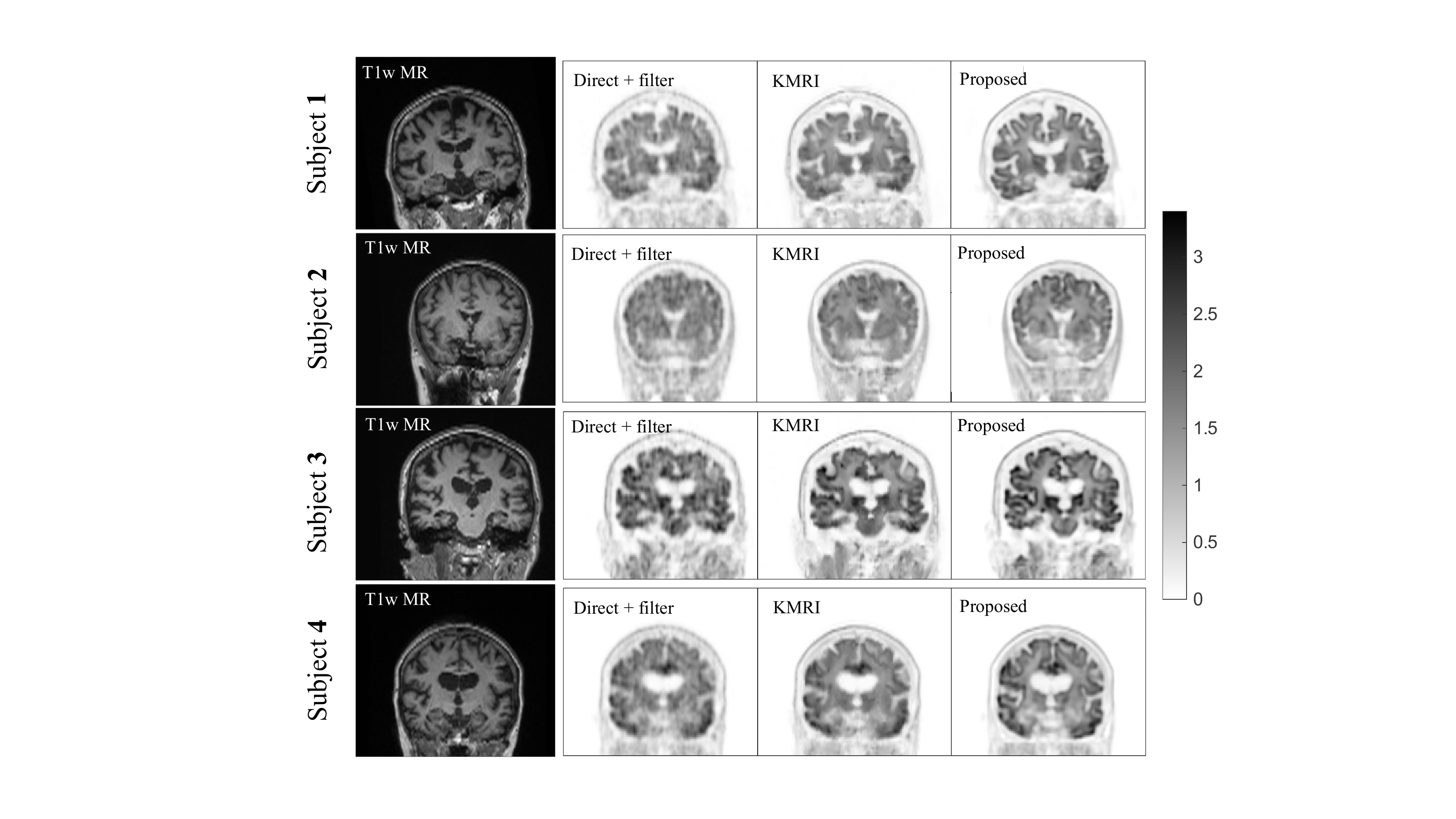}}
\caption{\small{ Coronal views of the DV images for different methods and different datasets. The different rows stand for the results of the four different datasets. The first column shows the corresponding T1-weighted MR prior image.} }
\label{fig:images_relogan}
\end{figure*}

\begin{figure}[t]
\centering
\subfloat[]{\includegraphics[width=1.7in]{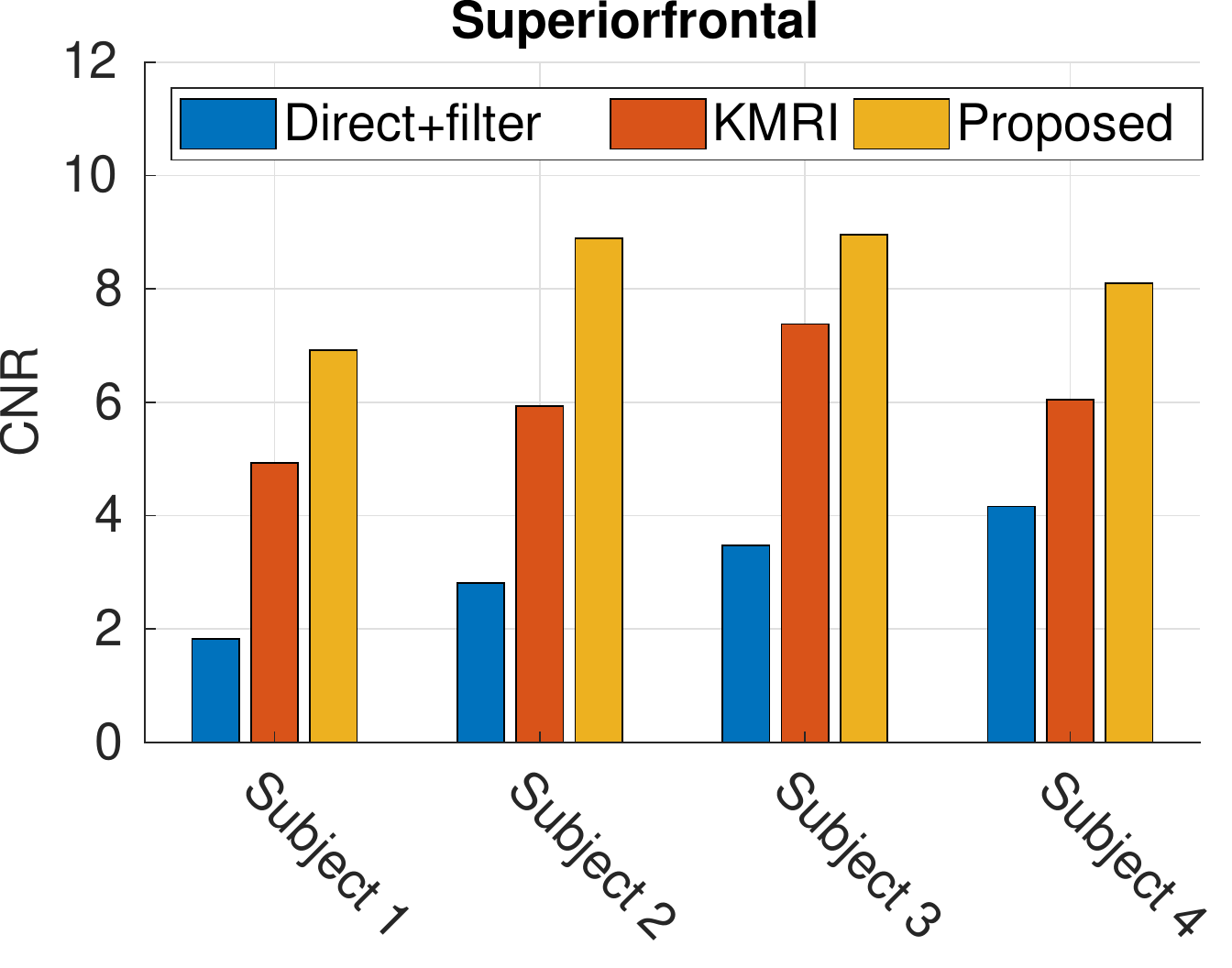}}
\subfloat[]{\includegraphics[width=1.7in]{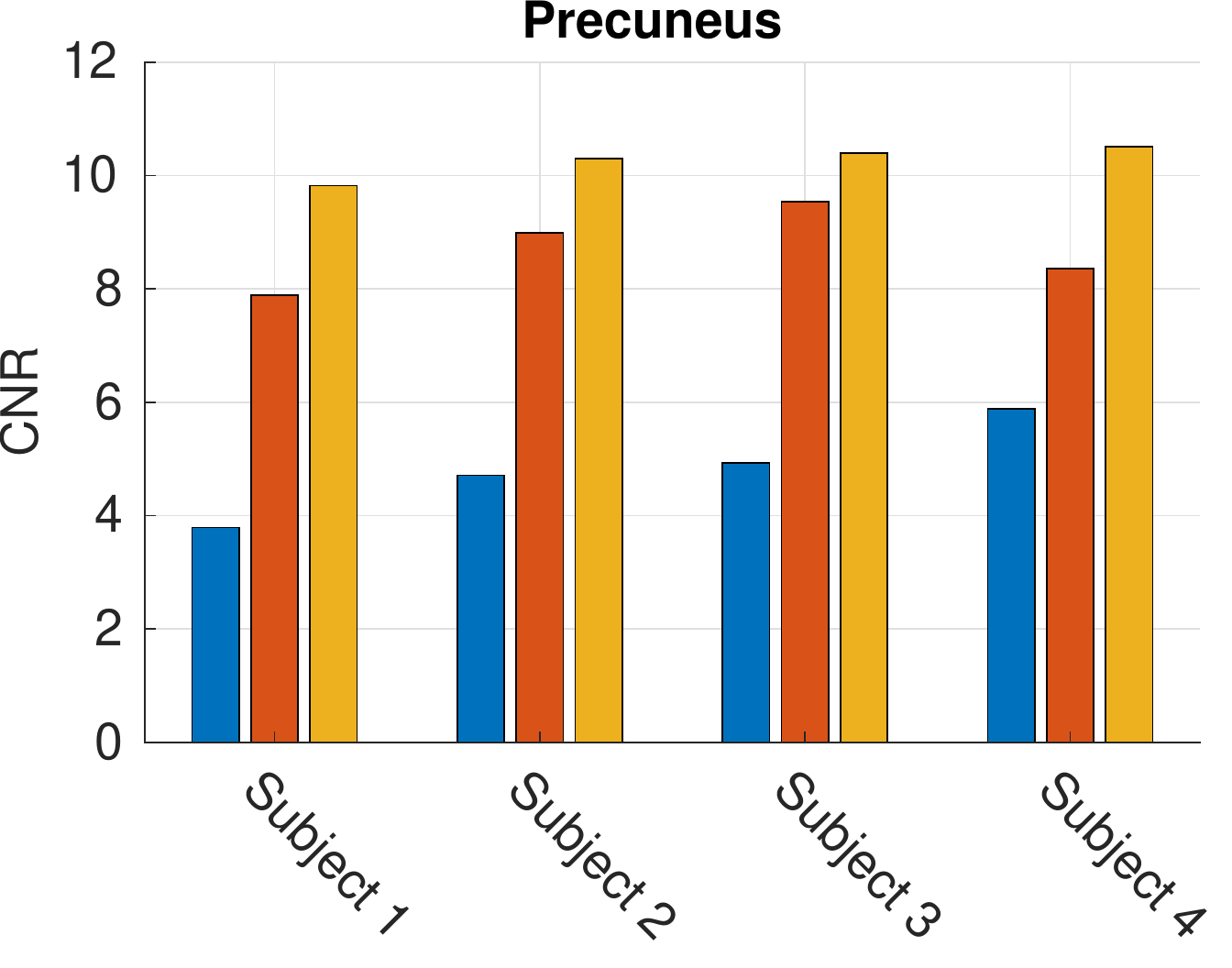}} \\
\subfloat[]{\includegraphics[width=1.7in]{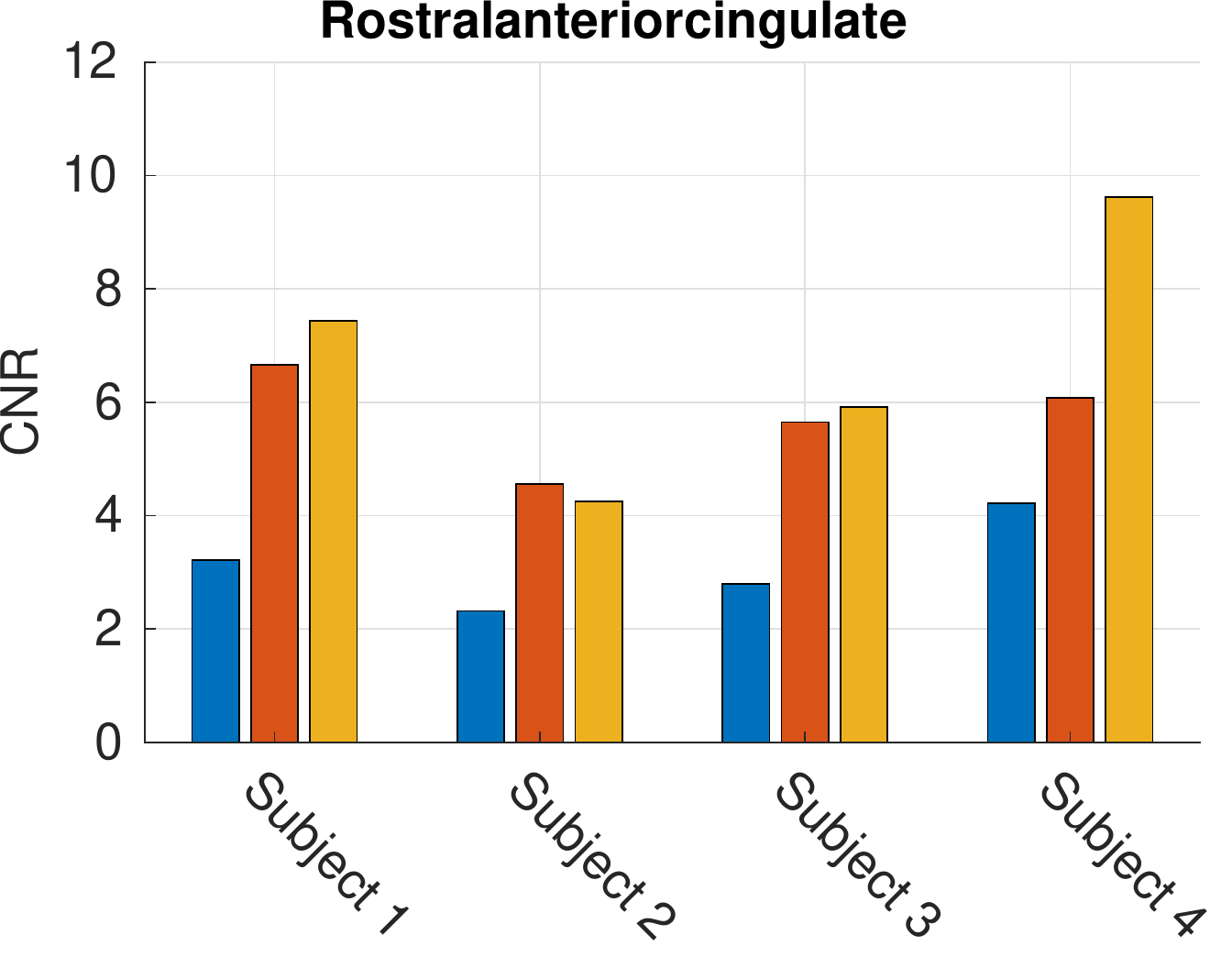}} 
\subfloat[]{\includegraphics[width=1.7in]{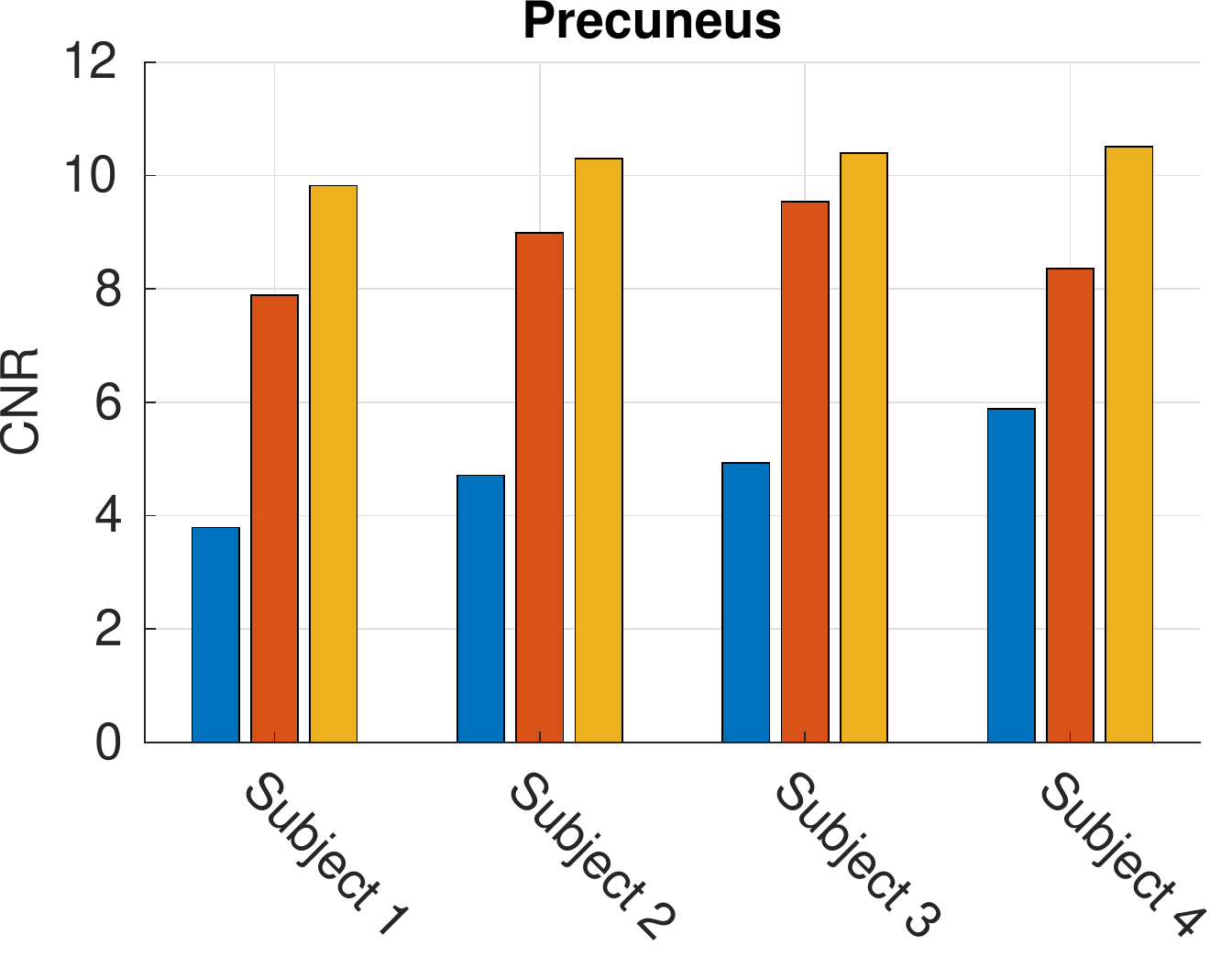}} \\
\subfloat[]{\includegraphics[width=1.7in]{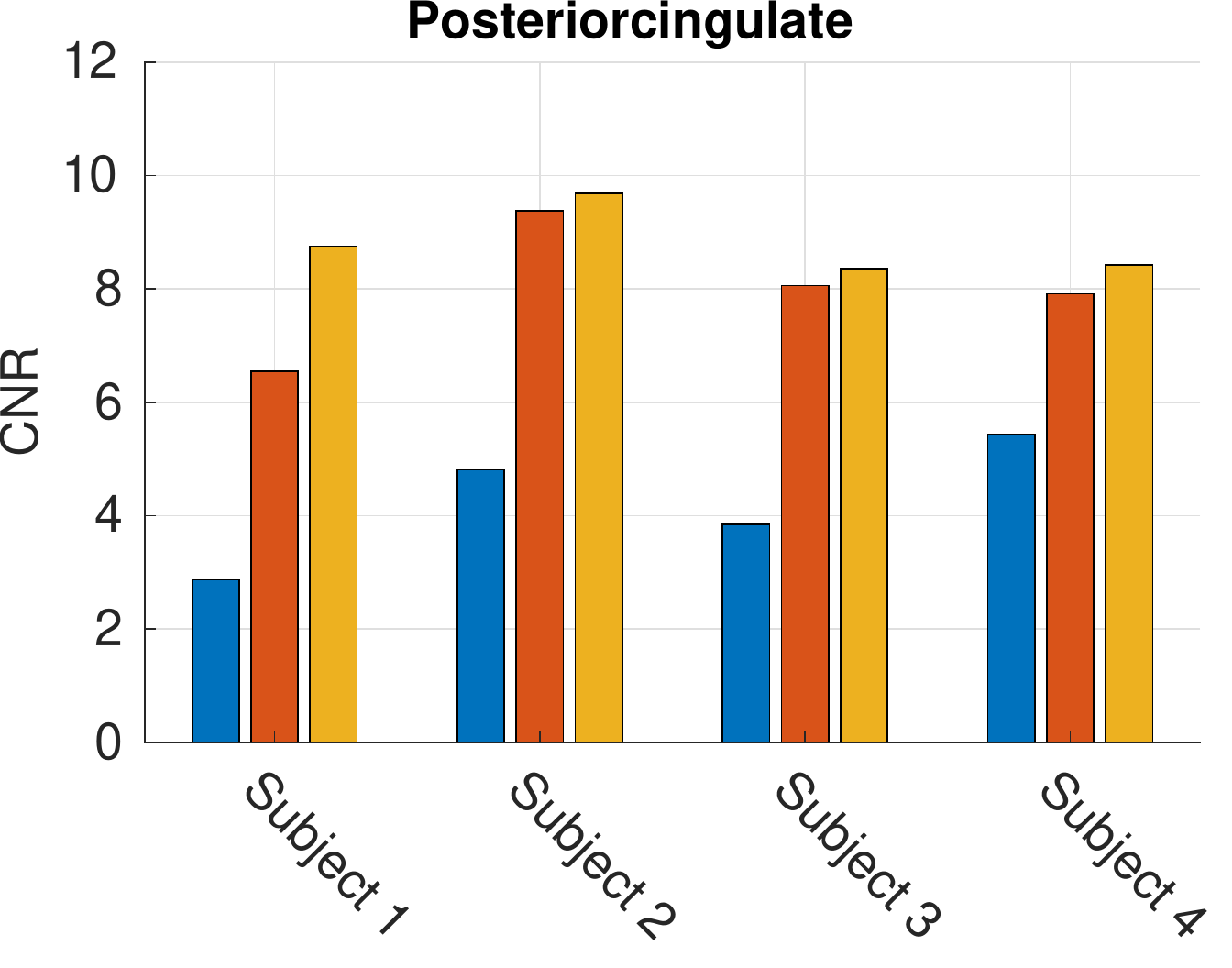}} 
\subfloat[]{\includegraphics[width=1.7in]{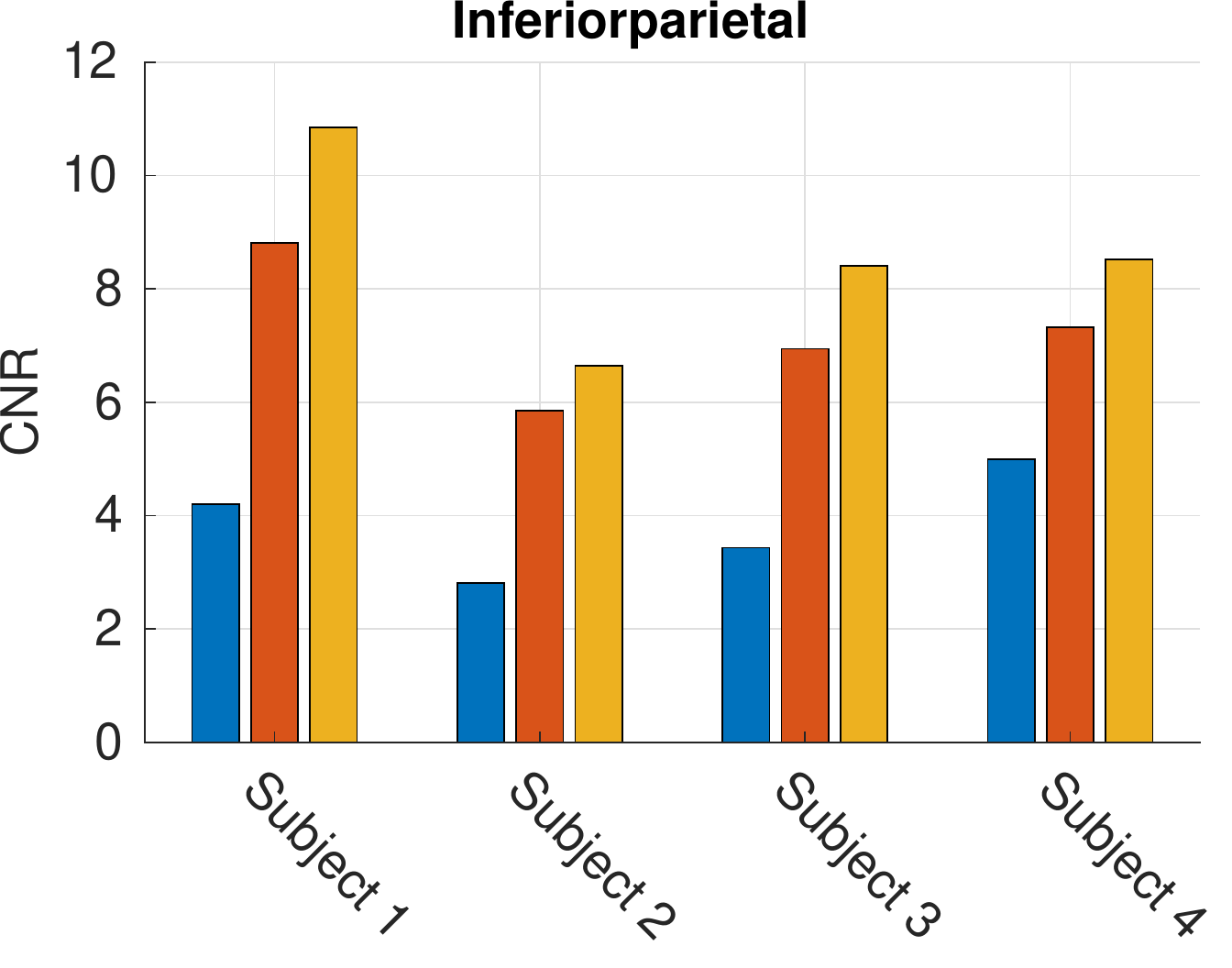}}
\caption{\small{ Quantification comparison of the CNR of different brain regions for the four C11-PiB datasets : (a) superfrontal, (b) supramarginal, (c) rostral anterior cingulate, (d) precuneus, (e) posterior cingulate, and (f) inferior parietal cortical regions. } }
\label{fig:logan_quantitative}
\end{figure}

\subsection{Real data for the Patlak model}
To validate the proposed method for the Patlak model, a 70-minutes low-dose dynamic 18F-FDG PET dataset with total counts equivalent to 1 mCi dose injection was used. The dataset was acquired from the Siemens Brain MR-PET scanner. The dynamic PET data was divided into 25 frames: 4$\times$20 s, 4$\times$40 s, 4$\times$60 s, 4$\times$180 s, 8$\times$300 s and 1$\times$600 s.  For quantitative comparison in the case where MRI and PET information does not match, an artificial spherical lesion of diameter 12.5 mm was inserted to the PET data (invisible in the MRI image). For the direct Patlak reconstruction, the last six frames were used ($ t^{\ast} = 30 $ min).  The data were reconstructed into an image array of 256$\times$256$\times$153 voxels with a voxel size of 1.25$\times$1.25$\times$1.25 $\text{mm}^3$. To obtain the blood input function, blood regions were segmented from a simultaneously acquired T1-weighted MRI image. Uptake in the inserted tumor and the left caudate region were measured. The image noise was calculated as the mean standard deviation of eleven circular background ROIs  (diameter = 12.5 mm, 10 pixels) from the white matter.

\subsection{Real data for the RE Logan model}\label{sec:pib_study}
To validate the proposed method for the RE Logan model, 60-minutes dynamic 11C-PIB PET scans of four mild cognitive impairment (MCI) patients were acquired on the GE DMI PET-CT scanner after 555 MBq bolus injection. T1-weighted anatomical images were acquired on the 3T Siemens MAGNETOM Trio MR scanner.  The dynamic PET data were divided into 39 frames: 8$\times$15 s, 4$\times$60 s, and 27$\times$120 s. The data were reconstructed into an image array of 256$\times$256$\times$89 voxels with a voxel size of 1.17$\times$1.17$\times$2.8 $\text{mm}^3$.  Fig.~\ref{fig:validation_relogan} shows that for the 11C-PiB tracer, the slope for the Logan and the RE Logan model is very close to each other for the precuneus and superior frontal cortices. Based on Fig.~\ref{fig:validation_relogan}, we have chosen the last 7 frames (44 min - 60 min, $ t_2^{\ast} = 46 $ min) for direct reconstruction. Rigid registration was performed using ANTs \cite{avants2011reproducible} to map the PET and MR images, as well as motion correction of the dynamic PET series. The motion transformation matrix was included in the direct image reconstruction for all methods. FreeSurfer \cite{fischl2004automatically} was used for MR parcellation to get brain ROIs. Cerebellum cortex was chosen as the reference region. Eleven circular regions (diameter = 11.7 mm) drawn from the white matter with approximately uniform uptakes were chosen as the background ROIs. Inferior parietal, precuneus, posterior cingulate, rostral anterior cingulate, superior frontal and supramarginal, the widely used ROIs for amyloid burden quantification \cite{johnson2016tau}, were chosen for the contrast-to-noise (CNR) calculation, which was defined as 
\begin{equation}
\text{CNR} = (\text{DV}_{\text{cortical}} - \text{DV}_{\text{back}})/ \text{STD},
\end{equation}
where $\text{DV}_{\text{brain}}$ is the DV value of the cortical ROI,   $\text{DV}_{\text{back}}$ is the mean DV value of the background ROIs, and $\text{STD}$ is the mean standard deviation of the background ROIs .
\section{Results}
\subsection{Simulation results}
We first tested the effectiveness of the proposed kernel layer by performing the network training using the network with and without the kernel layer. The training epoch is 1000 based on the L-BFGS optimizer. The results are shown in Fig.~\ref{fig:kernel_layer}. We can observe that the proposed kernel layer can further reduce the image noise while also better preserving the brain structures. Fig.~\ref{fig:simu_img_appear} shows three views of the Patlak-slope images reconstructed using different methods along with the ground-truth image. It can be observed that adding the anatomical prior information based on the kernel method and the proposed method can both reduce the image noise and better resolve the cortical details. Compared to the kernel method, the proposed method has better recoveries of the cortical details.  In addition, the shape of the inserted tumor regions, where there are mismatches between PET and MR prior images, were better preserved by the proposed method.  Fig.~\ref{fig:simu_quantitative} shows the quantification results of the gray matter region and the inserted tumor region for different methods at different iteration numbers. The proposed method has the best performance regarding the bias vs. noise trade-off.

\subsection{Real data results for the Patlak model}

Fig.~\ref{real_quantitative_patlak} shows three views of the reconstructed Patlak-slope images along with the MR prior image. The direct reconstruction results based on the $\text{EM} + \text{filter}$ baseline method are still noisy due to limited counts. Both the kernel method and the proposed method can improve the image quality by leveraging the high-quality MR prior image. The images obtained by the proposed method show the highest lesion contrast with clearer cortical structures as compared with other methods.  Fig.~\ref{real_quantitative_patlak} shows the uptake vs. noise curves for different methods at different iteration numbers. It can be observed that the proposed method has the best performance for both the left-caudate and tumor ROIs. 

\subsection{Real data results for the RE Logan model}

Fig.~\ref{fig:images_relogan} shows the coronal views of the DV images from four datasets for different methods. Compared to the $\text{EM} + \text{filter}$ baseline method, both the kernel method and the proposed method can improve the image quality by revealing more cortical details and reducing the image noise in the white matter. Fig.~\ref{fig:logan_quantitative} shows the CNR results for the four datasets of different cortical regions. Results show that the proposed method has the best performance for most cortical regions across the four subjects.

\section{Discussion}
For dynamic PET, it is difficult to obtain high-quality training labels, as the scanning time/injected dose is difficult to be further increased. Compared to static PET, more information exists in the noisy dynamic PET data itself. These two aspects make unsupervised deep learning more appealing for dynamic PET. In this work, we proposed an unsupervised deep learning framework for direct PET parametric image reconstruction. A new CNN was specifically designed to represent dynamic PET image series, with the same patient's high-quality prior image as the network input to provide a manifold constraint. Both the Patlak and RE Logan models were investigated in this work to demonstrate the feasibility of the proposed framework for irreversible and reversible tracers. Simulation and real data results show that the proposed framework can have better performance than other reference methods. It should be noted that the prior MR images needed for this framework can come from either a simultaneous PET/MR acquisition as presented in Fig.~\ref{real_quantitative_patlak}, or a stand-alone MR acquisition as shown in Fig.~\ref{fig:images_relogan}. 

As for the network structure,  3D Unet was adopted as the backbone in our work due to its strong representation power. To better utilize the anatomical prior information, an additional nonlocal operation based on the proposed kernel layer was embedded in the network to yield additional feature denoising. Results shown in Fig.~\ref{fig:kernel_layer} demonstrate the effectiveness of this nonlocal operation. This proposed kernel layer does not introduce additional training parameters and is computational efficient through the pre-calculation of the kernel matrix. Further developing more advanced network structures to enable better parametric generation is one of our future works. 

Furthermore, the Patlak and RE Logan models were embedded in the network graph as kinetic-model layers to generate the final dynamic PET image series based on the parametric images generated through the 3D Unet. For the RE Logan model, we proposed a new binning strategy and a constrained-optimization approach to preserve the i.i.d. assumption of the PET raw data. Though dynamic frames were thus coupled, the image reconstruction algorithm developed in this work based on the optimization transfer framework still enabled efficient frame-by-frame reconstruction. For other nonlinear kinetic models, such as the two-tissue compartment model (2TCM) and the simplified reference tissue model (SRTM), they can also be embedded into the network graph by defining the gradients with respect to each parametric parameter to enable back-propagation, which is one of our future works. 

\section{Conclusion}
In this work, we proposed a nonlocal deep image prior-based approach for direct parametric reconstruction based on the Patlak and the RE Logan model. The nonlocal operation was achieved by a kernel matrix layer and the kinetic model was embedded as a convolutional layer in the network. Computer simulation and real data evaluations demonstrate the effectiveness of the proposed method over other reference methods. Future work will focus on more quantitative evaluations.

\section{Acknowledgments}
The authors would like to thank Dr. Keith A. Johnson from MGH for sharing the 11C-PiB datasets. 
\end{twocolumn}


\end{document}